\documentclass[useAMS,usegraphicx,usenatbib]{mn2e}
\newcommand*{\Scale}[2][4]{\scalebox{#1}{$#2$}}%
\DeclareMathSymbol{\not}{\mathrel}{symbols}{"36}

\title[New chemical evolution analytical solutions]{New chemical evolution analytical solutions including environment effects}

\author[E.~Spitoni]{E.~Spitoni,$^{1}$\thanks{E-mail: spitoni@oats.inaf.it}
  \\
  $^1$ Dipartimento di Fisica, Sezione di Astronomia, Universit\`a di Trieste, via G.B. Tiepolo 11, I-34131, Trieste, Italy 
}
\usepackage[percent]{overpic}

\begin{document}
\date{Accepted . ; in original form xxxx}

\pagerange{\pageref{firstpage}--\pageref{lastpage}} \pubyear{xxxx}

\maketitle

\label{firstpage}

\begin{abstract}
  In the last years, more and more interest has been devoted to
  analytical solutions, including inflow and outflow, to study the
  metallicity enrichment in galaxies.  In this framework, we assume a
  star formation rate which follows a linear Schmidt law, and we
  present new analytical solutions for the evolution of the
  metallicity (Z) in galaxies. In particular, we take into account
  environmental effects including primordial and enriched gas infall,
  outflow, different star formation efficiencies and galactic
  fountains.  The enriched infall is included to take into account
  galaxy-galaxy interactions.  Our main results can be summarized as:
  i) when a linear Schmidt law of star formation is assumed, the
  resulting time evolution of the metallicity Z is the same either for
  a closed-box model or for an outflow model. ii) The mass-metallicity
  relation for galaxies which suffer a chemically enriched infall,
  originating from another evolved galaxy  with no pre-enriched gas, is shifted down in parallel
  at lower Z values, if compared the closed box model. iii) When a
  galaxy suffers at the same time a primordial infall and a chemically
  enriched one, the primordial infall always dominates the chemical
  evolution.  iv) We present new solutions for the metallicity
  evolution in a galaxy which suffers galactic fountains and an
  enriched infall from another galaxy at the same time.  The
  analytical solutions presented here can be very important to study
  the metallicity (oxygen), which is measured in high-redshift
  objects. These solutions can be very useful: a) in the context
  of cosmological semi-analytical models for galaxy formation and
  evolution, and b) for the study of compact groups of galaxies.
\end{abstract}

\begin{keywords}
galaxies: abundances - galaxies: evolution - galaxies: ISM
\end{keywords}

\section{Introduction}

 The galactic chemical evolution is the study of the transformation of
 gas into stars and the resulting evolution of the chemical
 composition of a galaxy. The so called ``Simple Model'' remains an
 useful guide for understanding the chemical evolution of galaxies
 since the pioneering works of Schmidt (1963), Searle \& Sargent
 (1972), Tinsley (1974), Pagel \& Patchett (1975).  In order to derive
 analytical solutions for the chemical evolution of galaxies one needs
 to make several hypotheses: the initial mass function (IMF) should be
 considered constant, the lifetime of stars should be neglected (
 instantaneous recycling approximation, IRA), and the complete mixing
 of chemical elements with the surrounding interstellar medium (ISM).
 These assumptions, in fact, allow us to have analytical expressions
 for the metallicity evolution of the galaxies in time and in terms of
 the gas fraction.  One can find analytical solutions even
 in presence of gas flows (infall, outflow, galactic fountains) as
 shown by Tinsley (1980), Clayton (1988), Lacey \& Fall (1985),
 Matteucci \& Chiosi (1983), Edmunds (1990), Recchi et al. (2008),
 Spitoni et al. (2010), Peeples \& Shankar (2011),  Lilly et
 al. (2013), Pipino et al. (2014), Peng \& Maiolino (2014a), Recchi \&
 Kroupa (2014), Kudritzki et al. (2015).
 
However, in all mentioned works the solutions are obtained only under
 specific assumptions regarding the infall/outflow rates. However, we
 note that analytical solutions are not able to give a complete
 description of the chemistry of a galaxy, since they fail in
 following the evolution of elements created on long time-scales, such
 as iron and nitrogen. A satisfactory description of the iron
 evolution requires detailed numerical models relaxing IRA and
 including the chemical enrichment from Type Ia SNe (e.g.
 Matteucci \& Greggio 1986) allow one to follow in detail the
 evolution of single elements. Matteucci \& Greggio (1986) showed in
 detail the effect of the time-delay model, already suggested by
 Tinsley (1980) and Greggio \& Renzini (1983), in particular the
 effect of a delayed Fe production by Type Ia SNe on abundance ratios
 involving $\alpha$-elements (O, Mg, Si). Analytical solution can be,
 on the other hand, adopted when studying the evolution of oxygen,
 created on short timescales and tracing the evolution of the
 global metallicity, Z, of which oxygen is the main component.

 In this work we start by using the formalism described by Matteucci
 (2001), Recchi et al. (2008), and Spitoni et al. (2010) and we show,
 for the first time, the analytical solution of the evolution of the
 metallicity of a galaxy in presence of ``environment'' effects
 coupled with galactic fountains and primordial infall of gas. In this
 context for ``environment'' effect we mean the situation where a
 galaxy suffers, during its evolution,  infall of enriched gas from
 another evolving galactic system, and this gas represents an enriched
 infall variable in time.

 The dynamics of interacting systems have been the subject of many
 papers concerning numerical simulations (Toomre \& Toomre 1972;
 Barnes \& Hernquist 1992; Berentzen et al. 2003) and
 spectrophotometric models (Larson \& Tinsley 1978; Kennicutt 1990;
 Temporin et al.  2003a, 2003b). Large amounts of ISM can be removed
 from the main disks of spiral galaxies by different processes: tides
 due to the gravitational force of a companion, ram pressure stripping
 during a near head-on collision between gas-rich galaxies, ram
 pressure stripping by intracluster gas, and galactic winds driven by
 supernovae.  Smith \& Struck (2001) using CO signatures observed 11
 extragalactic tails and bridges in nine interacting galaxy systems.
 Recently, Smith et al. (2010) using Galaxy Evolution Explorer (GALEX)
 ultraviolet telescope studied star formation morphology and stellar
 populations and in 42 nearby optically selected pre-merger
 interacting galaxy pairs.  Tails and bridges structures are often
 more prominent relative to the disks in UV images compared to optical
 maps. This effect is likely due to enhanced star formation in the
 tidal features compared to the disks rather than to reduced
 extinction. We also refer the reader to the review of Boselli \&
 Gavazzi (2006) where a comprehensive description of the environment
 effects on late-type galaxies in nearby clusters is presented. Among
 them we recall the tidal interactions among galaxy pairs which act on
 gas, dust and stars, as well as on dark matter and is depending on
 the gravitational bounding of the various components; tidal
 interaction between galaxies and the cluster potential well, and
 finally the so-called ``galaxy harassment'': the evolution of cluster
 galaxies is governed by the combined effect of multiple high speed
 galaxy-galaxy close ($\sim$ 50 kpc) encounters with the interaction
 with the potential well of the cluster as a whole.

 As stated by Davies et al. (2010), galaxy-galaxy interactions are
particularly important and striking when the speed of the interaction
is well matched to the velocities of the stars and gas. Therefore,
small galaxy groups can potentially provide the environment for
dramatic gravitational disturbances.  Beyond the Local Group, the
closest example of this is the environment around M81, the M81
group. In fact, in M81 group it is evident that various galaxies are
connected by flows.  Extended filamentary structures
external to the disc of M81, are clearly seen in emission in all of the
Herschel bands. These complex interactions cannot be described by the
simple tools offered, for example, by Recchi et al. (2008) or similar
papers in literature, and with this paper we provide a new set of
analytical solutions to take into account this kind of interactions in
the framework of analytical chemical evolution models.

Moreover, an enriched infall of gas can be originated from the galaxy
itself.  In fact, in the galactic fountain models (Shapiro \& Field
1976,  Houck \& Bregman 1990), hot gas is ejected out of
the Galactic disk by supernova (SN) explosions, and part of this gas
falls back in the form of condensed neutral clouds which move at
intermediate and high radial velocities. For example, in the galactic
fountain model the ejected gas from SN events falls back ballistically
(Bregman 1980).  These models are able to explain the vertical motion
of the cold and warm gas components observed in several spiral
galaxies (e.g. Fraternali et al. 2004; Boosma et al. 2005). Following
the analytical implementation of the galactic fountain presented by
Recchi et al. (2008), we will consider such an effect in our new
analytical solutions.

In the context of cosmological semi-analytical models of galaxy
formation and evolution each galaxy is treated as one unresolved
object, using integrated properties to describe the mass of stars,
cold gas, hot gas and the black hole. Since each component of the
galaxy is represented by one number, the dynamics within the galaxy is
not resolved, and one needs to assume with laws for star formation,
cooling and feedback that are valid on average for the entire
galaxy. In this context, our new analytical solutions shall be
extremely useful, because they give simple recipes concerning the time
evolution of the global metallicity of a galaxy in different
situations including inflow and outflow.

The paper is organized as follows: in Sect. 2 we present the main
assumptions of the ``analytical'' chemical evolution models, in
Sect. 3 we describe our assumptions and the system of equation we need
to solve, and in Sect 4 new analytical solutions are presented.
Finally, our conclusions are summarized in Section 5. In Appendix A we
draw the complete expressions of some new analytical solutions we
presented in this paper,   and in Appendix B a list of variables and parameters used throughout the paper is reported.
\section{The  Closed Box and leaky-box prescriptions }

The main assumptions of the Simple Model (Tinsley 1980) are:

\begin{enumerate}

\item The IMF is constant in time.

\item The gas is well mixed at any time ({\it instantaneous mixing approximation}).

\item Stars $\geq$ 1 M$_\odot$ die instantaneously; stars 
smaller than 1 M$_\odot$ live forever ({\it instantaneous recycling
approximation} or IRA).

\end{enumerate}

\noindent
These simplifying assumptions allow us to calculate analytically the
chemical evolution of the galaxies once we have defined the
fundamental quantities, such as the returned fraction:

\begin{equation}
R = \int_1^\infty (m - M_R) \phi (m) dm,
\label{eq:r}
\end{equation}
\noindent
(where $\phi (m)$ is the IMF and $M_R$ is the mass of the remnant) and the yield per 
stellar generation:

\begin{equation}
y_Z = {1 \over {1 - R}} \int_1^\infty m p_{Z, m} \phi (m) dm,
\label{eq:yield}
\end{equation}
\noindent
(where $p_{Z, m}$ is the fraction of newly produced and ejected metals
by a star of mass $m$).

Recently, Recchi \& Kroupa (2015) applied the integrated galactic
initial mass function (IGIMF) to the simple model solution, and $y_Z$
and $R$ are not constant, but are functions of time,through the time
dependence of the star formation rate (SFR) and the metallicity of the
system.

The well known solution of the so called ``closed-box''  where the system is one-zone and  there are no inflows
nor outflows  with constant  mass (gas plus stars)  is:

\begin{equation}
Z = y_Z \ln (\mu^{-1})
\label{eq:simple}
\end{equation}

\noindent
where $\mu$ is the gas fraction $M_{gas}/M_{tot}$, with $M_{tot}=M_* +
M_{gas}$.  It is also assumed that the $M_{gas}(0)=M_{tot}(0)$ and the initial metallicity of the system is zero.

Analytical solutions of simple models of chemical evolution including
infall or outflow are known since at least 30 years (Pagel \& Patchett
1975, Hartwick 1976, Clayton 1988, Twarog 1980, Edmunds 1990).  Here,
we follow the approach and the terminology of Matteucci (2001), where
it was assumed for simplicity {\it linear} flows ( gas flows are
proportional to the  SFR).  Therefore, the
outflow rate $W (t)$ is defined as:
 
\begin{equation}
W (t) = \lambda (1 - R) \psi (t),
\label{eq:w}
\end{equation}
\noindent
where $\psi (t)$ is the SFR, and the infall rate $A(t)$ is given by:

\begin{equation}
A (t) = \Lambda (1 - R) \psi (t).
\label{eq:a}
\end{equation} 
\noindent
Here $\lambda$ and $\Lambda$ are two proportionality constants $\ge
0$.  The first assumption is justified by the fact that, the larger
the SFR is, the more intense are the energetic events associated with
it (in particular supernova explosions and stellar winds) and
therefore the larger is the chance of having a large-scale outflow
(see e.g. Silk 2003).  A proportionality between $A (t)$ and $\psi
(t)$ has been discussed by Recchi et al. (2008). They tested different
prescriptions for the infall of gas showing that their results do not
change substantially if a generic exponential infall is assumed. This
demonstrates also that the major source of error in the solutions of
the simple models is the IRA assumption rather than the assumption of
linear flows.

 In Lilly
et al (2013) and Pipino et al. (2014), they considered the simple
chemical evolution model in the cosmological context where the gas
accreted is proportional to the dark matter growth. In particular, in
Pipino et al. (2014) the infall parameter $\Lambda(t)$ is time dependent
 and  is defined as the ratio between the accretion 
rate from cosmological simulations and the SFR.  In this paper we do
not investigate ``cosmological'' aspects but we study  the effects
of galaxy-galaxy interactions on the time evolution of galactic
metallicity.

The evolution of the metallicity of a system as a function of $\mu$  in the case with only outflows, i.e.
$A(t) = 0 $ and $W(t) \not= 0$ is the following:
\begin{equation}
Z = {{y_Z} \over (1+\lambda)}\ln [(1+\lambda)\mu^{-1}-\lambda],
\label{zout}
\end{equation}
with the assumption that at $t=0$, $Z$(0)=0,
$M_{tot}(0)=M_{gas}(0)$. 
In the opposite case ($A(t) \not= 0 $ and $W(t)=0$), 
assuming for the infalling gas a primordial composition  (e.g $Z_A=0$), we obtain this solution (Matteucci 2001):
\begin{equation}
Z = {{ y_Z} \over \Lambda}\biggl\lbrace 
1 - \bigl[ (\Lambda- (\Lambda  - 1) \mu^{-1} \bigr]^{\Lambda \over {\Lambda   
- 1}} \biggr\rbrace.
\label{zinfall}
\end{equation}
\noindent
The general solution presented by Matteucci (2001) for a system described by the simple model in
the presence of infall of gas with a general metallicity $Z_A$ and
outflow is:
\begin{equation}
Z = {{\Lambda Z_A + y_Z} \over \Lambda}\biggl\lbrace 
1 - \bigl[ (\Lambda - \lambda) 
- (\Lambda - \lambda - 1) \mu^{-1} \bigr]^{\Lambda \over {\Lambda - \lambda 
- 1}} \biggr\rbrace.
\label{eq:sol}
\end{equation}

 \section{The chemical evolution model with environment effects} Using
the formalism introduced by Matteucci (2001), and adopted later by
Recchi et al. (2008) and Spitoni et al. (2010), for the first time we
intend to study the effects of the environment on the chemical
evolution of a galaxy by means of analytical solutions. The
environment effects are mimicked by an infall of gas with a time
dependent metallicity originated by a  nearby galaxy.

In our model we analyze the evolution of the oxygen in a generic galaxy, and
assume for the oxygen yield standard  values of $y_O$ = 0.01 and the
returned fraction  $R=0.25$. As done in Spitoni et al. (2010), those
values were obtained by adopting the Salpeter (1955) IMF, and the
stellar yields of Woosley \& Weaver (1995) for oxygen at solar
metallicity.

\subsection{Our model for a system formed by two isolated galaxies}

We consider the chemical evolution of the galaxy1 with initial mass
$M_{tot1}(0)= M_{g1}(0)$.   We  study here the effects of  the interaction
with another galaxy  on the abundance of oxygen
$Z_{O,1}(t)=M_{O,1}(t)/M_{g1}(t)$.

We assume an environment dependent infall, originated by feedback
  episodes and galaxy-galaxy interactions. We
  assume that the infall is proportional to the gas outflows from
  nearby galaxies with wind parameter $\lambda$. First we consider the
   case with  the ``enriched infall'' proportional to the SFR of the galaxy2:

\begin{equation}
  A_{\lambda,1} (t) =\epsilon  W_{\lambda,2} (t)=\epsilon \lambda (1 - R) \psi_2 (t),
\end{equation}

  where $\psi_2 (t)$ is the SFR of the galaxy2, and $\epsilon$ is the
fraction of outflowing gas from galaxy2 which reaches galaxy1. We
assume in this work that $\epsilon=0.5$. In fact, the outflows are
generally bipolar and for a lobe approaching galaxy 1, the second lobe
moves away from it.  The value $\epsilon$=0.5 is an upper limit
for the fraction of outflowing gas from galaxy2 to galaxy1. Only if
the outflow is well collimated along a narrow solid angle and only if
galaxy1 happens to intercept this galactic wind, $\epsilon$ can be
equal to 0.5. For example, from GALEX UV images of the starbust galaxy
M81 given by Hoopes et al. (2005), it can be seen that the opening
angle of the northern side wind is around 55 degrees.

  The $Z_{O,2}(t)$  metallicity of the infall   is  time
  and depends on the outflows of the  galaxy2.

  The system of equations we have to solve for our galaxy1 is the
following :

\begin{equation}
\cases{\Scale[1.3]{{d M_{tot1} \over d t}} = (\epsilon \lambda) (1 - R) \psi_2 (t) \cr
\Scale[1.3]{{d M_{g1} \over d t}} =  (1 - R)\big(\epsilon\lambda  \psi_2 (t)-  \psi_1 (t)\big) \cr
\Scale[1.3]{{d M_{O,1} \over d t}} = (1 - R) \big(\psi_1 (t) \big[ y_O - Z_{O,1}(t)\big]+  \psi_2 (t) \epsilon \lambda Z_{O,2}(t)\big),}
\label{system1}
\end{equation}
where $\psi_1 (t)$ is the SFR of the galaxy1.

As mentioned above, $Z_{O,2}(t)$ is the metallicity of the galaxy2 which is 
suffering only gas outflows.  We therefore need to solve another
system of equations for the evolution of  $Z_{O,2}(t)$ in galaxy2.

Recalling Matteucci (2001) the system to be solved for galaxy2 with
only outflow of gas is:

\begin{equation}
\cases{\Scale[1.3]{{d M_{tot2} \over d t}} = -\lambda (1 - R) \psi_2 (t) \cr
\Scale[1.3]{{d M_{g2} \over d t}} = (-\lambda  -1) (1 - R) \psi_2 (t) \cr
\Scale[1.3]{{d M_{O,2} \over d t}} = (1 - R) \psi_2 (t) \big[ -Z_{O,2}(t)+ y_O - \lambda Z_{O,2}(t)\big].}
\label{system2}
\end{equation}
We recall that the outflow rate for the galaxy2 is $W _{\lambda,2}(t)
=\frac{A_{\lambda,2} (t)}{\epsilon}$.  With equations (\ref{system1})
and (\ref{system2}) we study an isolated system formed by two galaxies
when $\epsilon=1$. In fact, in this case we have that ${d
M_{tot2} \over d t}+{d M_{tot1} \over d t} = 0$.

On the other hand, imposing $\epsilon \not= 1$, we assume that part
of the chemical enriched gas escaping from  galaxy2, ends up in
the in intergalactic medium (IGM).

\subsection{Model including primordial infall of gas  and the interactions of 2 galaxies}

The majority of detailed chemical evolution models of galactic systems
assumes that the galaxies formed by accretion of primordial gas from
the IGM.  Therefore, we examine here also the more realistic case
where both galaxy1 and galaxy2 suffer an inflow of primordial gas. We
assume that the infall rate is
proportional of the SFR of the galactic system as discussed in Section
2.1. Hence, for the galaxy1 the primordial infall rate is
$A_{\Lambda,1} (t)=\Lambda (1 - R) \psi_1 (t)$, whereas is
$A_{\Lambda,2} (t)=\Lambda (1 - R) \psi_2 (t)$ for the galaxy2.  The
systems of equation we need to solve are the following
ones: \begin{itemize}
\item galaxy1:
\begin{equation}
\cases{\Scale[1.3]{{d M_{tot1} \over d t}} =  (1 - R)(\epsilon \lambda \psi_2 (t) + \Lambda \psi_1 (t)) \cr
\Scale[1.3]{{d M_{g1} \over d t}} =  (1 - R)\big(\epsilon\lambda  \psi_2 (t)+(\Lambda-1)  \psi_1 (t)\big) \cr
\Scale[1.3]{{d M_{O,1} \over d t}} = (1 - R) \big(\psi_1 (t) \big[ y_O - Z_{O,1}(t)\big]+  \psi_2 (t) \epsilon \lambda Z_{O,2}(t)\big).}
\label{sprimordial1}
\end{equation}

\item galaxy2:

\begin{equation}
\cases{\Scale[1.3]{{d M_{tot2} \over d t}} = (-\lambda +\Lambda)(1 - R) \psi_2 (t) \cr
\Scale[1.3]{{d M_{g2} \over d t}} = (\Lambda-\lambda  -1) (1 - R) \psi_2 (t) \cr
\Scale[1.3]{{d M_{O,2} \over d t}} = (1 - R) \psi_2 (t) \big[ -Z_{O,2}(t)+ y_O - \lambda Z_{O,2}(t)\big].}
\label{sprimordial2}
\end{equation}

\end{itemize}

\section{The new analytical solutions}

In this Section we present our new analytical solutions. First, we
 discuss the case where the galaxy1 suffers an enriched infall from
 an evolving companion galaxy (galaxy2). In the following we
 generalize this solution in the case of different SFEs for the
 galaxy1 and galaxy2. We also present the case where a primordial infall
 is taken into account. The last result is related to
 ``environment'' effects coupled with galactic fountains.

\subsection{Interactions with a nearby galaxy}
From the system of equations (\ref{system2})  we recover $Z_{O,2}(t)$, and following Matteucci (2001) we obtain:

\begin{equation}
Z_{O,2}(t)=\frac{y_O}{1+\lambda} \ln \left[(1+\lambda)\mu^{-1}_2- \lambda\right].
\label{z2}
\end{equation}
Assuming a linear Schmidt (1959) law ($\psi= S \times M_g$) (Pipino et al. 2014,  Recchi et al, 2008), in the systems (\ref{system1}) and (\ref{system2}), we
have  the following expressions of the time evolution  of the gas for the
galaxy1 ($M_{g1}(t)$) and the galaxy2 ($M_{g2}(t)$):

 \begin{equation}
 M_{g1}(t) = e^{-(1 - R)St}\big(M_{g1}(0)+ \epsilon M_{g2}(0)\big[1- e^{-\lambda(1 - R)St}\big]\big),
\label{Mg1_SFE_fix}
\end{equation}

\begin{equation}
 M_{g2}(t) =  M_{g2}(0)e^{-(\lambda  +1) (1 - R)St}.
\end{equation}
Recalling that the gas fraction is defined as $\mu=M_{g}(t)/M_{tot}(t)$, we can write for the galaxy2:
\begin{figure}
	  \centering   
    \includegraphics[scale=0.42]{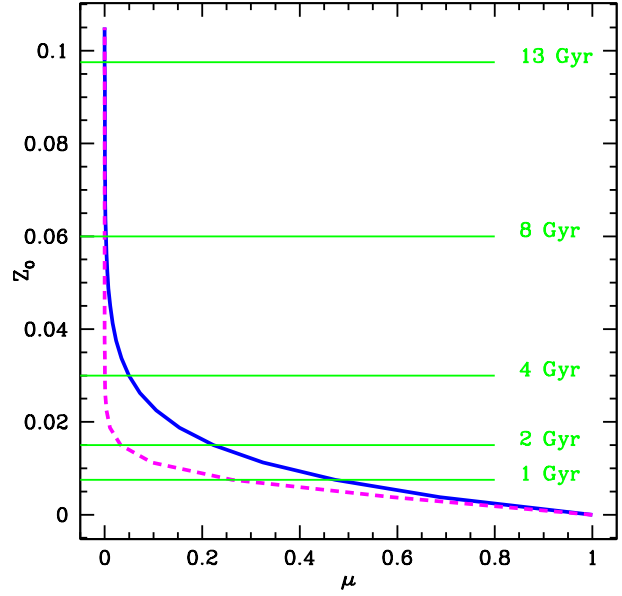} 
   \caption{ Evolution of the metallicity $Z_O$ in terms of the gas mass
     fraction $\mu$  for the closed-box model (blue solid line) and for
     a model with only gas outflow with $\lambda=2$ (magenta dotted line). Different horizontal solid green lines refer to different galactic time (1, 2, 4, 8,  and 13 Gyr).  }
		\label{wt}
\end{figure} 

\begin{equation}
\mu^{-1}_2=\frac{M_{tot2}}{M_{g2}}=\frac{M_{tot2}}{M_{g2}(0)e^{-(\lambda  +1) (1 - R)St}}.
\label{mumeno}
\end{equation}
From the system (\ref{system2}) with the initial condition that
$M_{tot2}(0)=M_{g2}(0)$, we have that:
\begin{equation}
M_{tot2}(t)=\frac{\lambda}{\lambda+1}M_{g2}(t)+M_{g2}(0)\left(1-\frac{\lambda}{\lambda+1}\right).
\label{mtot2}
\end{equation}
Finally, inserting  eq. (\ref{mtot2}) in eq. (\ref{mumeno}), we recover the following time dependent expression for $\mu^{-1}_2(t)$: 
\begin{equation}
\mu^{-1}_2=\Scale[1.1]{\frac{\lambda}{\lambda+1}+e^{(\lambda  +1) (1 - R)St}\left(\frac{1}{\lambda+1}\right)}.
\end{equation}
Hence, the expression of $Z_{O,2}(t)$  as a function of the galactic time becomes:

\begin{equation}
Z_{O,2}(t)=\frac{y_O}{1+\lambda} \ln \left[\lambda+e^{(\lambda  +1) (1 - R)St}- \lambda\right]=y_O(1 - R)St.
\label{z2}
\end{equation}

It is worth noting that the evolution in time of $Z_{O,2}(t)$ does not
depend on $\lambda$ in the specific case of the Schmidt (1959) SFR
with $k$=1, and is the same as in the closed-box one.  In fact, for a
closed-box model with the linear Schmidt (1959) law  we
have this expression for the gas mass: $ M_{g}(t) = M_{g}(0)e^{-(1 - R)St}$.  Following
eq. (\ref{eq:simple}) we derive  the expression of the metallicity
$Z_{cb}$ as a function of time for the closed-box ($cb$). Therefore, we  have  that:
\begin{equation}
\begin{array} {lcl} 
Z_{cb}(t)=y_O\ln\left(e^{(1 - R)St}\right)=y_O(1 - R)St.
\end{array}
\label{cb}
\end{equation}

In Fig. \ref{wt} we draw the evolution of the metallicity
$Z_O$ as  function of  the gas fraction $\mu$ for both closed-box model and the one
with linear outflows in the case of the  linear Schmidt (1959) law,
$\psi=SM_{g}$, with wind parameter $\lambda$=2 and $S$=1 Gyr$^{-1}$.  Horizontal lines indicate
different galactic times during the galactic histories.
\begin{figure}
	  \centering \includegraphics[scale=0.42]{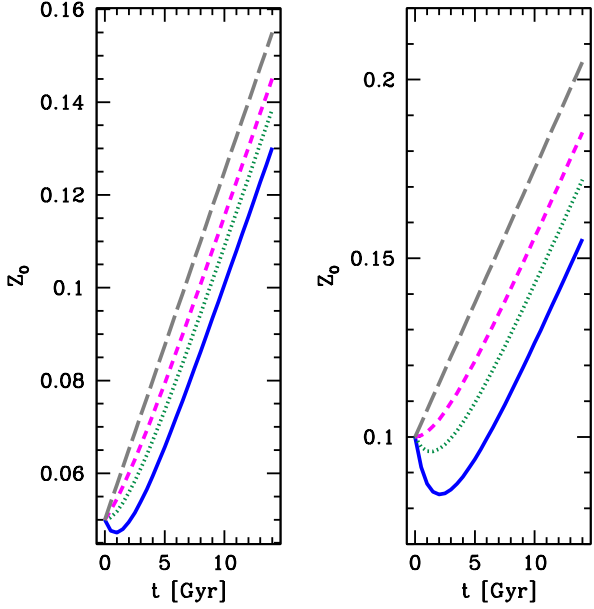} \caption{{\it
    Left panel:} The time evolution of the metallicity (oxygen) for
    the galaxy1 when galaxy-galaxy interactions are taken into account
    and assuming an initial metallicities fixed at the value of
    $Z_{O,1}(0)$=0.05  and $Z_{O,2}(0)$=0. For all the cases we
    assume, $\lambda$=0.4, $\epsilon$=0.5, S=1 Gyr$^{-1}$.
      The magenta short dashed line represents the case
    with $M_{g1}(0)/ M_{g2}(0)$=2, the green dotted line for
    $M_{g1}(0)/ M_{g2}(0)$=1, the solid blue line represents the case
    with $M_{g1}(0)/ M_{g2}(0)$=0.5. The long dashed grey line is the
    closed-box solution with the same initial metallicity. {\it Right
    panel:}: Same of left panel but with an initial metallicity fixed
    at $Z_{O,1}(0)$=0.1.  } \label{window}
\end{figure}

\begin{figure}
\centering \includegraphics[scale=0.42]{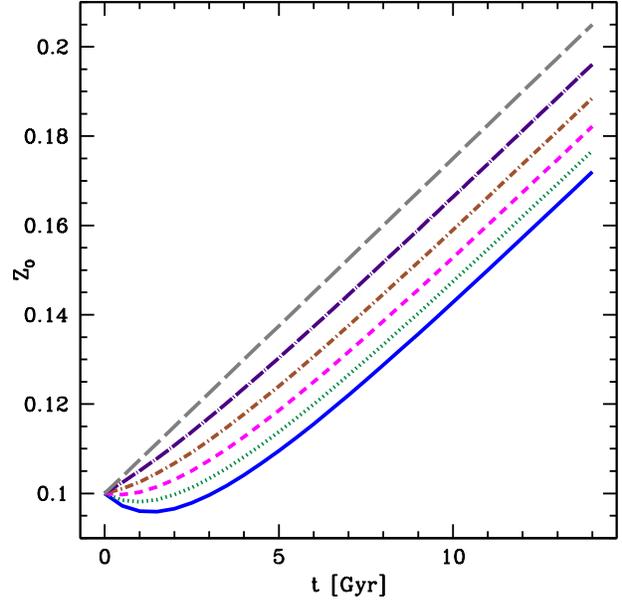} 
\caption{ The effect of the fraction of the outflowing gas  ($\epsilon$) from galaxy2 to galaxy1  on the time evolution of the metallicity (oxygen) for
    the galaxy1.   We assume $Z_{O,1}(0)$=0.1, $Z_{O,2}(0)$=0,
    $\lambda$=0.4, S=1 Gyr$^{-1}$, and $M_{g1}(0)/M_{g2}(0)$=1.  The
    blue solid line represents the model with $\epsilon$=0.5, the case
    with $\epsilon$=0.4 is reported with the dotted green line, the
    short dashed magenta line is the case with $\epsilon$=0.3. The
    models with $\epsilon$=0.2 and $\epsilon$=0.1 are in brown dashed
    dotted line and violet long dashed dotted line, respectively.  The
    long dashed grey line is the closed-box solution with the same
    initial metallicity $Z_{O,1}(0)$=0.1.}
\label{epsilon}
\end{figure}

Even if  the closed-box model and the
model with only outflow have the same metallicity in time, that same
metallicity is reached for different values of the gas
fraction $\mu$.

The metallicity $Z_{O,2}(t)$ expressed in eq. (\ref{z2}) can be inserted in the system (\ref{system1}), and recalling that $\frac{d M_{O,1}}{dt}$=$\frac{d (M_{g1}Z_{O,1}(t)) }{dt}$, the third equation of the system (\ref{system1}) can be rewritten as:
\begin{displaymath}
M_{g1}(t)\frac{d Z_{O,1}(t)}{dt} = (1 - R)  \times
\end{displaymath}
\begin{equation}
\big(\psi_1 (t) y_O + \epsilon  \psi_2 (t) \big[-\lambda Z_{O,1}(t) +\lambda Z_{O,2}(t)   \big]\big)
) . 
\label{ets}
\end{equation}
Hence, we assume that the SFR follows a Schmidt (1959) law: $\psi_1= S \times M_{g1}$ and $\psi_2= S \times M_{g2}$, and
 the differential equation we need to solve when we include eq. (\ref{z2}) in eq. (\ref{ets})  is:

\begin{displaymath}
\frac{d Z_{O,1}(t)}{dt} = (1 - R) S \times
\end{displaymath}
\begin{equation}
\Big(  y_O +   \frac{M_{g2}(t)}{M_{g1}(t)} \big[-\epsilon \lambda Z_{O,1}(t) +\epsilon \lambda y_O (1-R)St   \big] \Big).
\label{nonho}
\end{equation}

The final expression for $Z_{O,1}(t)$ in a case where the galaxy1 is
affected by the interaction with another galactic system (galaxy2),
and with $Z_{O,2}(0)=0$ is the following one:
\begin{equation}
Z_{O,1}(t) = Sy_O(1-R)t+ \frac{ Z_{O,1}(0)}{1+  \epsilon \frac{M_{g2}(0)}{M_{g1}(0)}\Big(1 - e^{-\lambda (1 - R)St}\Big) }. 
\label{znew}
\end{equation}

We see that the closed-box solution is still recovered when the
 initial metallicity $Z_{O,1}(0)$ for the galaxy1 is considered equal
 to zero.   The reason for this behaviour is that the outflow from
 galaxy2 has the same metallicity of the whole galaxy, i.e. it is
 expressed by eq. (20). Since the two expressions (eqs. 20 and 21) are
 identical, and since metallicity is not an additive quantitiy, the
 expression $Z_{O,1}(t)$ in eq. (24) is expected.   In
 Fig. \ref{window} we show the time evolution of the metallicity
 (oxygen) where we assume an initial metallicity different from zero
 in the galaxy1 and as function of different $M_{g1}(0)/ M_{g2}(0)$
 ratios: 2, 1 and 0.5.  We considering two different initial
 metallicities $Z_{O,1}(0)$=0.05 (left panel) and $Z_{O,1}(0)$=0.1
 (right panel). We are aware that these values are extremely large,
 but here we only like to show  the trends of considering
 different pre-enriched values.  We see that at early times the
 dilution effect is prominent because the infalling gas has a much
 lower metallicity of that galaxy1, and as expected the lower the
 ratio $M_{g1}(0)/ M_{g2}(0)$ is, more the dilution effect is
 important.

 In Fig.  \ref{epsilon} the effects of the fraction of outflowing
gas $\epsilon$ on the time evolution of the galaxy1 oxygen abundance
obtained by eq. (\ref{znew}) are shown.
Assuming  $Z_{O,1}(t)$=0.1,  $M_{g1}(0)/ M_{g2}(0)$=1, $\lambda$=0.4, and $S$= 1 Gyr$^{-1}$,  we present  different models varying $\epsilon$: $\epsilon= 0.5,
0.4, 0.3, 0.2, 0.1$. As expected, the dilution effect is larger
assuming higher $\epsilon$ values.

\begin{figure}
	  \centering \includegraphics[scale=0.42]{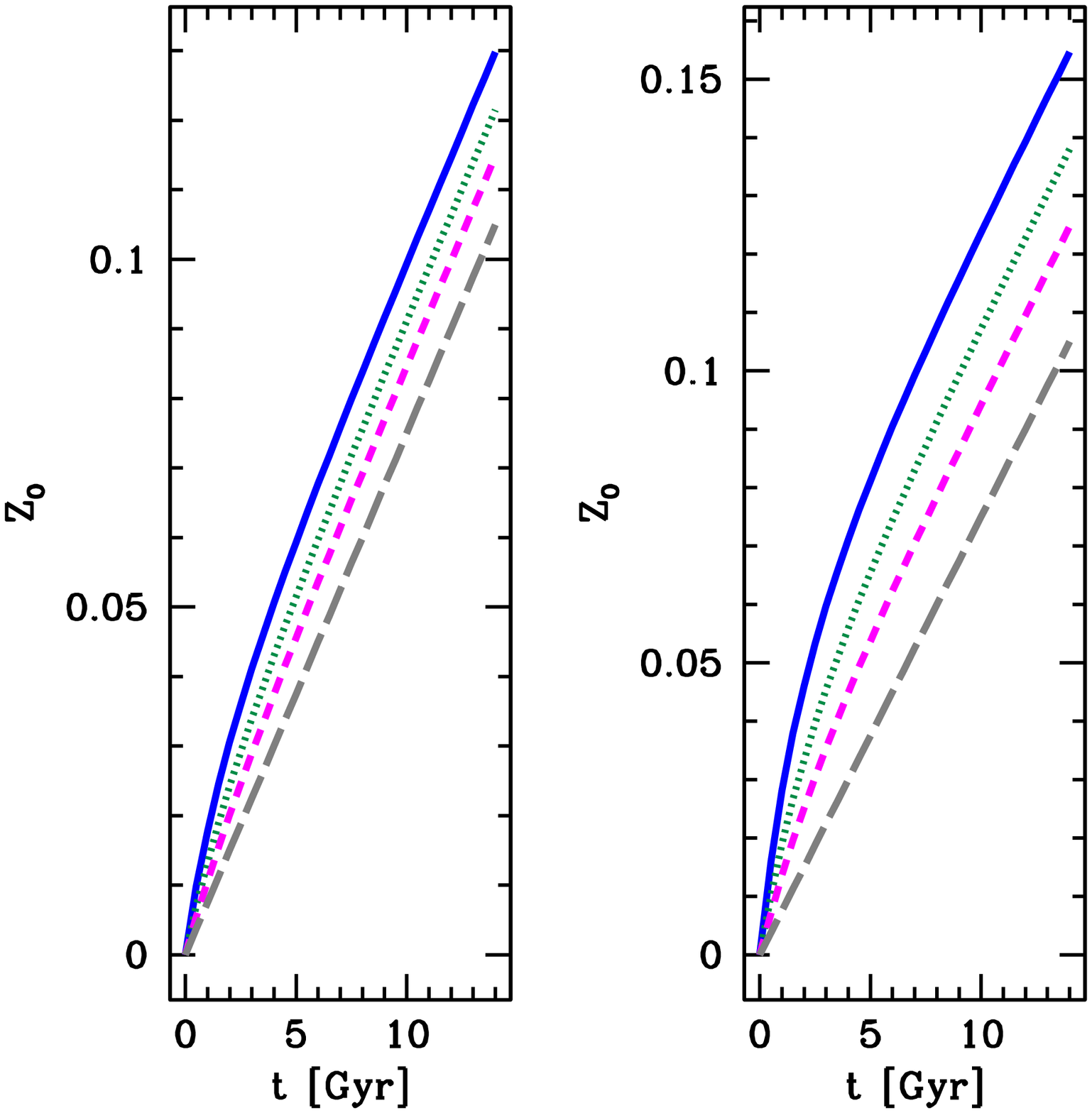} \caption{
 {\it
    Left panel:} The time evolution of the metallicity (oxygen) for
    the galaxy1 when galaxy-galaxy interactions are taken into account
    and assuming an initial metallicities fixed at the values of
    $Z_{O,1}(0)$=0 and    $Z_{O,2}(0)$=0.05 . For all the cases we assume, $\lambda$=0.4,
    $\epsilon$=0.5, S=1 Gyr$^{-1}$.  The magenta short
    dashed line represents the case with $M_{g1}(0)/ M_{g2}(0)$=2, the
    green dotted line for $M_{g1}(0)/ M_{g2}(0)$=1, the solid blue line
    represents the case with $M_{g1}(0)/ M_{g2}(0)$=0.5. The long
    dashed grey line is the closed-box solution with the same initial
    metallicity. {\it Right panel:}: Same of left panel but with the
    initial metallicity  fo galaxy2 fixed at $Z_{O,2}(t)$=0.1.  } \label{window_Z2}
\end{figure}

We discuss now the case where the initial metallicity for galaxy1 and galaxy2 are $Z_{O,1}(0)$=0,  $Z_{O,2}(0) \not=$ 0, respectively.
In this case the differential equation we need to solve  is:

\begin{displaymath}
\frac{d Z_{O,1}(t)}{dt} = (1 - R) S \times
\end{displaymath}
\begin{equation}
\Big(  y_O +   \frac{M_{g2}(t)}{M_{g1}(t)} \Big[-\epsilon \lambda Z_{O,1}(t) +\epsilon \lambda \Big(Z_{O,2}(0)+ y_O (1-R)St\Big)   \Big] \Big).
\label{nonho}
\end{equation}
 The new solution is:

\begin{equation}
Z_{O,1}(t) = Sy_O(1-R)t+ \frac{  Z_{O,2}(0)\epsilon \frac{M_{g2}(0)}{M_{g1}(0)}\Big(1 - e^{-\lambda (1-R)St}\Big) }{1+  \epsilon \frac{M_{g2}(0)}{M_{g1}(0)}\Big(1 - e^{-\lambda (1 - R)St}\Big)}. 
\label{znew_z2}
\end{equation}
In Fig. \ref {window_Z2} we show the evolution of the oxygen
abundance of galaxy1 using the new analytical solution presented in
eq. (\ref{znew_z2}) with $\lambda$=0.4, $\epsilon$=0.5, and
$Z_{O,1}(0)$=0, considering two different initial metallicities
$Z_{O,2}(0)$=0.05 and $Z_{O,2}(0)$=0.1. In this case we use such high
 values for the pre-enrichment in order to better visualize the
general trends and the effects of different $M_{g1}(0)/ M_{g2}(0)$
ratios. As expected, at variance with Fig. \ref {window}, the smaller
is the $M_{g1}(0)/ M_{g2}(0)$ ratios, the less efficient the chemical
evolution is for galaxy1.

Overall, the general solution for the evolution of the oxygen abundance for the
galaxy1 with both  initial metallicities $Z_{O,1}(0)$, $Z_{O,2}(0)$
different from zero, is given by the following expression:

\begin{equation}
Z_{O,1}(t) = Sy_O(1-R)t+ 
\label{znewgen}
\end{equation}
\begin{displaymath}
+\frac{ Z_{O,1}(0)+ Z_{O,2}(0)\epsilon \frac{M_{g2}(0)}{M_{g1}(0)}\Big(1 - e^{-\lambda (1-R)St}\Big) }{1+  \epsilon \frac{M_{g2}(0)}{M_{g1}(0)}\Big(1 - e^{-\lambda (1 - R)St}\Big) }. 
\end{displaymath}
 The general solution reported in eq. (\ref{znewgen}) describes the
more realistic scenario where galaxies exhibit different chemical
enrichment rates and therefore they posses different metallicity when
the interaction starts.   It seems plausible  that galaxy2
presents a faster chemical enrichment in systems as the M81 group,
where  large and metal rich galaxies (M81, M82, and NGC3077)
interact and eject gas. In other environments, it can be more
reasonable that only dwarf galaxies would show galactic winds, and therefore
in this situation galaxy2 is less chemically enriched than galaxy1. In
conclusion, both cases $Z_{O,2}(0)$$>$$Z_{O,1}(0)$,  and $Z_{O,2}(0)$$<$$Z_{O,1}(0)$
 have physical meaning.  Recently, in
Recchi et al. (2015) it was shown that the values of initial
metallicities depend on the level of the interaction between galaxies.

\subsection{Some applications to real cases}
The evolution of the stellar mass content
in the galaxy1 $M_{*,1}(t)$, can  be  simply inferred by the relation
$M_{*,1}(t)=M_{tot1}(t)-M_{g1}(t)$. The time evolution of the total
mass of galaxy1 $M_{tot1}(t)$, is given by  the first
equation of system (\ref{system1}), and with the initial condition
$M_{tot1}(0)=M_{g1}(0)$ we have that:

\begin{equation}
M_{tot1}(t)=M_{g1}(0)+\frac{\lambda \epsilon M_{g2}(0)}{1+\lambda}\big(1- e^{-\lambda(1 - R)St}\big).
\end{equation}

\begin{figure}
	  \centering \includegraphics[scale=0.42]{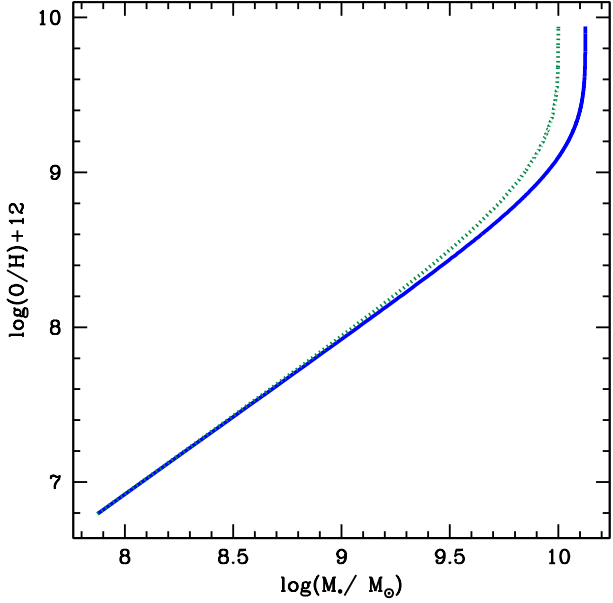} 
\caption{The
 evolution in the framework of the closed-box model of the abundance $\log (O/H)+12$ as a function of the stellar
 mass $M_*$ for the closed-box model for a galaxy with total mass
 equal to 10$^{10}$ $M_{\odot}$ is drawn with the dotted green
 line. With the solid blue line we show the results for the galaxy1
 with a initial mass of 10$^{10}$ $M_{\odot}$, and when we take into
 account the enriched infall from a companion galaxy with the same initial
 mass and the wind parameter equal to  $\lambda$=2, and $\epsilon$=0.5.  The initial oxygen abundances are  $Z_{O,1}(0)$=$Z_{O,2}(0)$=0.}   \label{ev_mstar}
\end{figure}

In Fig. \ref{ev_mstar} we compare the time evolution of
 log($O$/$H$)+12 as function of stellar mass for the closed-box model
 with a model where environment effects are taken into account.
 The log($O$/$H$)+12 quantity is recovered from $Z_{O,1}(t)$  using the following expression:
log$\Big(Z_{O,1}(t)/(16 \times 0.75)\Big)$+12, where 0.75 is the assumed fraction of hydrogen. 
  The
 assumed initial mass in each model is $10^{10}$ M$_{\odot}$.  In the last case, at a fixed stellar
 mass value the system shows a smaller 
 oxygen abundance compared to the closed-box evolution.

\begin{figure} 
	  \centering \includegraphics[scale=0.42]{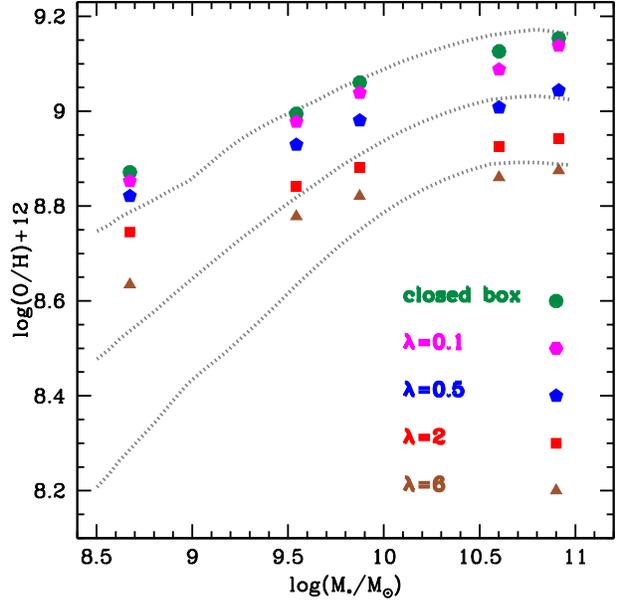} \caption{The
 observed mass-metallicity relation and related standard deviation of
 Kewley \& Ellison (2008) are indicated with the dotted grey
 lines. With the green circles we show the closed-box results for
 galaxies with total masses: 8.8$\times$10$^{8}$, 5$\times$10$^{9}$, 10$^{10}$,
 5$\times$10$^{10}$, 10$^{11}$ $M_{\odot}$, respectively. We compare them
 with model results where we consider the interactions with a
 companion galaxy with the same initial masses.  For all the models it is assumed that the initial oxygen abundances are  $Z_{O,1}(0)$=$Z_{O,2}(0)$=0. With the magenta
 hexagon models with the wind parameter fixed at the value of
 $\lambda=0.1$ are presented. The cases with $\lambda=0.5$,
 $\lambda=2$, and $\lambda=6$ are represented with blue pentagons, red
 squares, and brown triangles respectively.  } \label{MZ}
\end{figure}

In Fig. \ref{MZ} we study the effects of the environment on the
 mass-metallicity (MZ) relation by comparing with the observed
 relation of Kewley \& Ellison (2008) for Sloan Digit Sky Survey
 (SDSS) star-forming galaxies.  To estimate the amount of gas that
 resides in each star-forming galaxy, and the gas fraction $\mu$ as a
 function of the galactic stellar mass, we use the method described in
 Spitoni et al.  (2010).

We determine the cold gas mass of each galaxy
 on the basis of its SFR, using the  following  inverted  Kennicutt (1998) relation,
 which links the gas surface density to the SFR per unit area:

\begin{equation}
 \Sigma_{gas} = \big( \frac{ \dot{\Sigma}_{*} }{ 2.5 \times 10^{-4}}\big)^{0.714} \, \, \, \, \,    M_{\odot} \, pc^{-2}.
\end{equation}
where the gas density $\Sigma_{gas}$ is expressed in M$_{\odot} \, $pc$^{-2}$, and the SFR surface density $\dot{\Sigma}_*$  in M$_{\odot}
\, $yr$^{-1} \, $kpc$^{-2}$. 
The gas mass $M_{gas}$ (in $M_{\odot}$) is given by:
\begin{equation}
 M_{gas} = \Sigma_{gas} \times 2 \pi  R_{d}^{2}.
\end{equation}
where $ R_d$ is the scaling radius  calculated as in Mo
et al. (1998).
At this point we have a relation between $\mu$ and the stellar mass
for each considered galaxy.

 First, we follow the evolution of the closed box models with zero
 initial metallicity for different initial masses
 $M_{tot}(0)$=$M_{g}(0)$=$M_{tot}(t)$=8.8$\times$10$^{8}$,
 5$\times$10$^{9}$, 10$^{10}$, 5$\times$10$^{10}$, and 10$^{11}$
 M$_{\odot}$, respectively.  For each model we compute the time
 evolution of $Z_{O}(t)$, $\mu(t)$, and $M_{*}(t)$.  Thus, we consider
 the time $t_{MZ}$ where the stellar mass $M_*(t_{MZ}$) and the
 corresponding $\mu$ belong to the fit reported in Fig. 2 of Spitoni
 et al. (2010). At this point the metallicity at the time $t_{MZ}$ is
 computed using eq. (21).  In Fig. \ref{MZ} we show the closed-box
 model results for systems with initial masses $M_{tot}(0)=M_{g}(0)=
 M_{tot}(t)$ = 8.8$\times$10$^{8}$, 5$\times$10$^{9}$, 10$^{10}$,
 5$\times$10$^{10}$, and 10$^{11}$ M$_{\odot}$, respectively.

\begin{figure} 
	  \centering \includegraphics[scale=0.42]{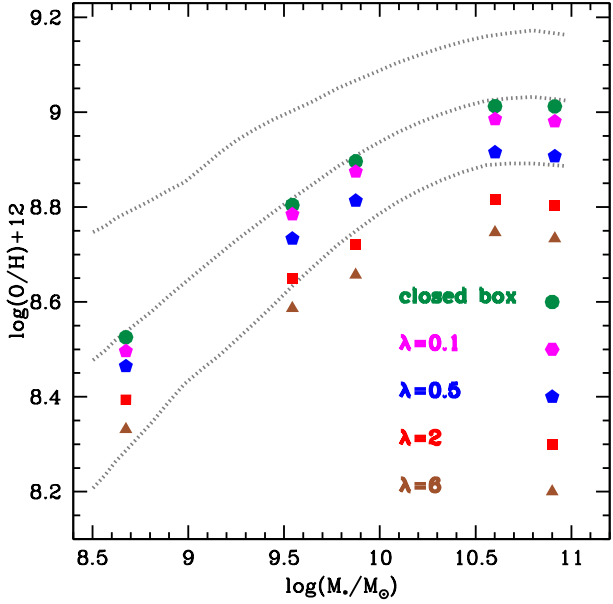} \caption{
 The observed mass-metallicity relation and related standard deviation
 of Kewley \& Ellison (2008) are indicated with the dotted grey
 lines. With the green circles we show the closed-box with the
 variable $y_O$ values of Spitoni et al. (2010) for galaxies with
 total masses as in Fig. \ref{MZ}. We compare them with model results
 where we consider the interactions with a companion galaxy with the
 same initial masses.  For all the models it is assumed that the
 initial oxygen abundances are $Z_{O,1}(0)$=$Z_{O,2}(0)$=0.  Models
 with different wind parameters $\lambda$ are indicated with symbols
 as Fig. \ref{MZ}. }  \label{MZ_yo}
\end{figure}

 The closed box model is not able to reproduce the observed MZ
 relation, and in Spitoni et al. (2010) we concluded that a galactic
 wind rate increasing with decreasing galactic mass or a variable IMF
 are both viable solutions for reproducing the MZ relation. We
 consider also galactic systems with the same initial masses of the
 closed-box models, but taking into account environment effects: an
 enriched infall from a galaxy with the same mass and with the same SFE   with zero initial metallicities for galaxy1 and galaxy2: $Z_{O,1}(0)$=$Z_{O,2}(0)$=0.

\begin{figure} 
	  \centering \includegraphics[scale=0.42]{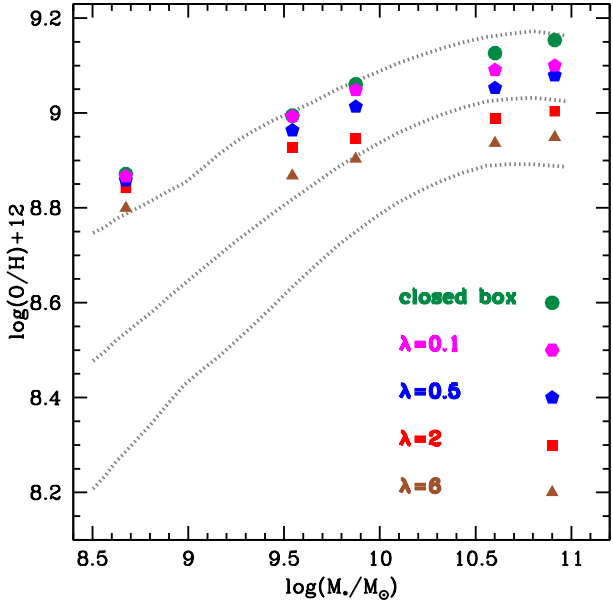} \caption{The
 observed mass-metallicity relation and related standard deviation of
 Kewley \& Ellison (2008) are indicated with the dotted grey lines.
 Models are indicated with symbols as Fig. \ref{MZ} assuming that the
 initial oxygen abundances are  for galaxy 1  $Z_{O,1}(0)$=0, and  for galaxy2  $Z_{O,2}(0)$= $5 \times10^{-3}$, respectively. } \label{MZ_Z2}
\end{figure} 

In Fig. \ref{MZ} we show the results when we consider four different
 inflow parameters: $\lambda$=0.1, 0.5, 2, 6.  We see that at a fixed
 stellar mass the metallicity of the galactic system drops down in
 presence of inflow. We are in agreement with the work of Torrey et
 al. (2012), where the effects of dynamical interaction of companion
 galaxies were studied.  In Spitoni et al. (2010) we have already
 proved that a combination of primordial infall and outflows with
 variable wind parameters as functions of the stellar mass is
 required to reproduce it. Here, we show the effects of this
 interaction compared to the closed-box case.

In
 Torrey et al. (2012) galaxy-galaxy interactions do not change the slope of the MZ, but the MZ
 is just shifted down in parallel at smaller metallicities. In our case, if we
 consider a constant $\lambda$  (in Fig. \ref{MZ} connecting the same
 type and colour points) for all the galaxy masses we have a
 flattening of the MZ slope compared to the closed-box. However, from
 Spitoni et al. (2010) we know that more massive galactic systems are associated 
 to smaller outflow episodes and therefore smaller  wind parameters. Hence, if we consider a variable
 $\lambda$ (decreasing towards higher mass galaxies) we preserve the
 slope of the MZ relation.

Our results are also in agreement with the observed depression of 0.05
-0.10 dex found by Ellison et al. (2008) for the MZ relation in
interacting galaxies, if we consider a variable wind parameter
$\lambda$ with a maximum value smaller than $\lambda$=6.

\begin{figure}
	  \centering \includegraphics[scale=0.42]{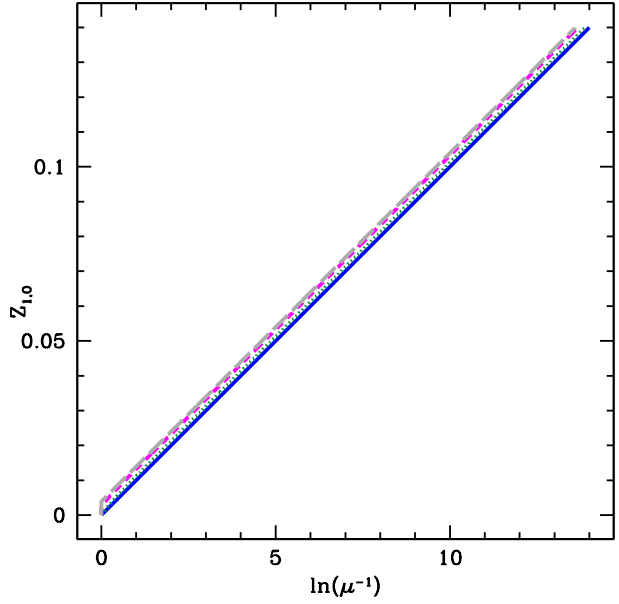} \caption{
 We show the metallicity $Z_{O,1}$ in the case of enriched infall from the
 evolution of galaxy2 as a function of $\ln(\mu^{-1})$. We fix
 $\epsilon$=0.5, $\lambda$=2, and consider different ratio $M_{g1}(0)/
 M_{g2}(0)$ values: 1 (green dotted line), 0.1
 (magenta short dashed line), 0.01 (grey long dashed line). We compare
 our results with closed model case (solid blue line).   The initial oxygen abundances for the galaxy1 and galaxy2 are  $Z_{O,1}(0)$=$Z_{O,2}(0)$=0. 
 } \label{logmu}
\end{figure} 

 In Fig.  \ref{MZ_yo} we show the MZ relation obtained by the
closed-box models with the same prescriptions adopted in Fig. \ref{MZ}
but with the variable yields $y_O$\footnote{We remind that varying the
yield for stellar generation corresponds to vary either the stellar
nucleosynthesis or the IMF.}  of Spitoni et al. (2010) which is able
to perfectly reproduce the MZ relation. Again, the fact of considering
``environment'' effects leads to a drop in the metallicity of the
galaxy1 at a fixed stellar mass compared to the closed-box models. The
fit with the observations deviates for $\lambda > 0.1$,
and this might be a hint that the kind of interactions treated in this
paper cannot be very general, but it surely applies for specific
systems (such as the M81 group).

Finally, in Fig.  \ref{MZ_Z2} we show the effects of different initial
metallicities $Z_{O,2}(0)$ of galaxy2 on the MZ relation. Assuming the
same model parameters of Fig. \ref{MZ} but with $Z_{O,2}(0)=
5 \times10^{-3}$, we see that at a fixed stellar mass, the decrease of
the metallicity is less prominent compared Fig. \ref{MZ} especially
for small stellar masses.

 Edmunds (1990) studied the effects of gas flows on the chemical
evolution of galaxies showing that it does exist a forbidden area in
the plane log($\mu^{-1}$) versus metallicity above the simple model
solution, for models with inflow of unenriched gas or outflows.  Here,
we want to test if this result is still valid in the case of a time
dependent enriched infall from a companion galaxy  with initial
metallicities for galaxy1 and galaxy2 set to zero: $Z_{O,2}(0)=Z_{O,1}(0)$=0.
The expression of $\mu_1(t)$ as a function of time is:
\begin{equation}
\mu_1(t)=\frac{ e^{-(1 - R)St}\big(1+ \epsilon \frac{M_{g2}(0)}{M_{g1}(0)}\big[1- e^{-\lambda(1 - R)St}\big]\big)}{1+\frac{\lambda \epsilon}{1+\lambda}\frac{M_{g2}(0)}{M_{g1}(0)}  \big(1- e^{-\lambda(1 - R)St}\big)}.
\label{mu1}
\end{equation}
We  need an expression for $\mu_1(t)$ as a function of the metallicity $Z_O$.
We have seen that the solution $Z_{O,1}(t)$ in the case of zero initial metallicity is $(1-R)Sy_ot$. Therefore, we can rewrite  eq. (\ref{mu1}) as:

\begin{equation}
\mu_1(Z_1)=\frac{ e^{-Z_1/y_O}\big(1+ \epsilon \frac{M_{g2}(0)}{M_{g1}(0)}\big[1- e^{-\lambda Z_1/y_O}\big]\big)}{1+\frac{\lambda \epsilon}{1+\lambda}\frac{M_{g2}(0)}{M_{g1}(0)}  \big(1-  e^{-\lambda Z_1/y_O}\big)}.
\end{equation}

In Fig. \ref{logmu} we show the metallicity $Z_{O,1}$ in case of
enriched infall from the evolution of galaxy2 as the function of
$\ln(\mu^{-1})$. We fix $\epsilon$=0.5, $\lambda$=2, and consider
different ratio $M_{g1}(0)/ M_{g2}(0)$ values:1, 0.1, 0.01. We see
that all the studied cases are in the forbidden area defined by
Edmunds (1990), and they evolve similarly to the closed box model with small differences.  This result, at variance with Edmunds
(1990) is due to the fact that here is the first time in which these
particular cases are studied and the theorems T(2) and T(3) of Edmunds
(1990) cannot be applied because of the presence of enriched infall.

\subsection{The analytical solution including  both galactic fountain and environment effects}
In Recchi et al. (2008), new analytical solutions in the framework of
differential winds were presented, namely galactic winds in which the
metals are ejected out of the parent galaxy more efficiently than the
other elements. The existence of differential winds has been first
introduced in the context of chemical evolution of galaxies by
Pilyugin (1993) and Marconi et al. (1994).  A realistic case of
variable infall metallicity is represented by the situation in which
the metallicity of the infalling gas is set to be always equal to the
one of the galactic wind.  This condition implies that the very same
gas that has been driven out of the galaxy by energetic events can
subsequently fall back to the galaxy, due to the gravitational
potential well.

\begin{figure}
	  \centering \includegraphics[scale=0.25]{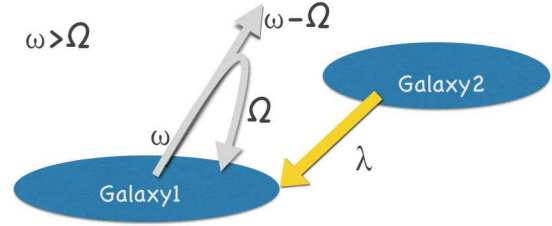} \caption{In
    this sketch the gas flows related the galaxy1 are represented
    when galactic fountain are considered. The enriched infall with
    the parameter $\lambda$ is drawn with the yellow arrow. The
    outflowing gas from galaxy1 with parameter $\omega$ and the
    fraction which fall back ($\Omega$) are in gray arrows. Our model
    is in the framework of differential wind theory when metals are
    more easily channeled out. }
\label{fontfig}
\end{figure}
\begin{figure}
	  \centering \includegraphics[scale=0.42]{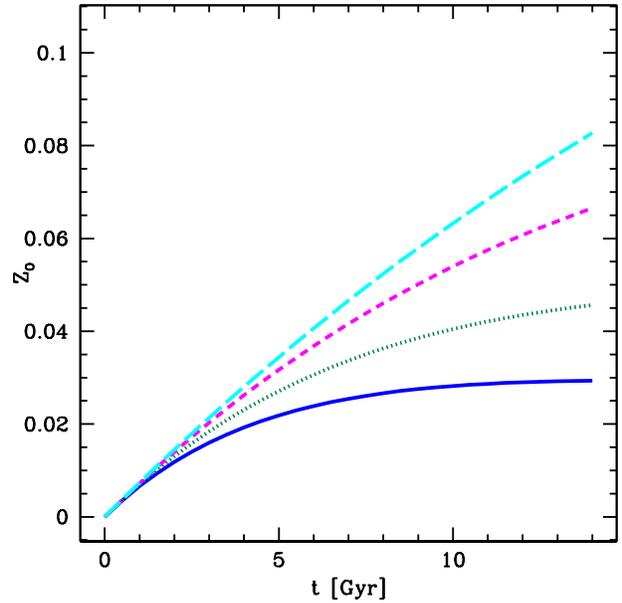} 
\caption{ We show the time evolution of the metallicity $Z_{O,1}$ in
    presence of galactic fountains when mass ratios $M_{g1}(0)/
    M_{g2}(0)=1$ and SFE $S$=1 Gyr$^{-1}$, using the new analytical
    solution presented in eq. (\ref{z_fountain}) with $\epsilon$=0.5,
    and $\lambda$=0.4, $\Omega=0.1$ and $\omega=0.2$.  The initial
    oxygen abundances for the galaxy1 and galaxy2 are
    $Z_{O,1}(0)$=$Z_{O,2}(0)$=0. The model with $\alpha$= 4.5 is drawn
    with the blue long solid line, the model with $\alpha$= 3 is
    represented by dotted green line solid line, the case with
    $\alpha$= 2 is drawn with with the short dashed magenta line. With
    the cyan long dashed line the model with $\alpha$= 1.5 is
    represented.}
\label{plotfont}
\end{figure}

\begin{figure}
	  \centering \includegraphics[scale=0.42]{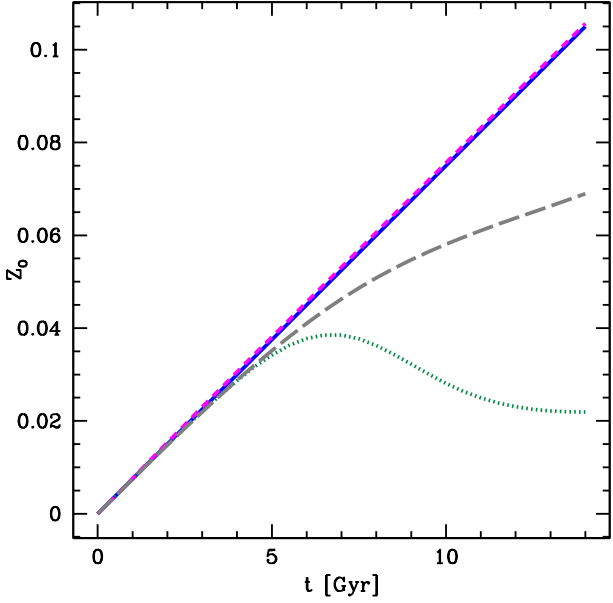} \caption{
 We show the time evolution of the metallicity $Z_{O,1}$ when we
 consider different SFEs $S_2$ for the galaxy2 following the new
 analytical solution presented in eq. (\ref{sol_SFE_dif}). We consider
 the case with $M_{g1}(0)/ M_{g2}(0)$=1,  $\epsilon$=0.5, and
 $\lambda$=0.1. The SFE for the galaxy1 is fixed at the value of
 $S_1$=1 Gyr$^{-1}$ .  The model with $S_2$=1
 Gyr$^{-1}$ is drawn with the blue solid line, the model with $S_2$=2
 Gyr$^{-1}$ is represented by the magenta short dashed line, With the
 long dashed grey line the model with $S_2$=0.5 Gyr$^{-1}$ is
 represented. Finally, the model with $S_2$=0.1 Gyr$^{-1}$ is drawn
 with the dotted green line. The initial oxygen abundances for the galaxy1 and galaxy2 are  $Z_{O,1}(0)$=$Z_{O,2}(0)$=0.}
		\label{1SFEV}
\end{figure}

\begin{figure}
	  \centering   
    \includegraphics[scale=0.42]{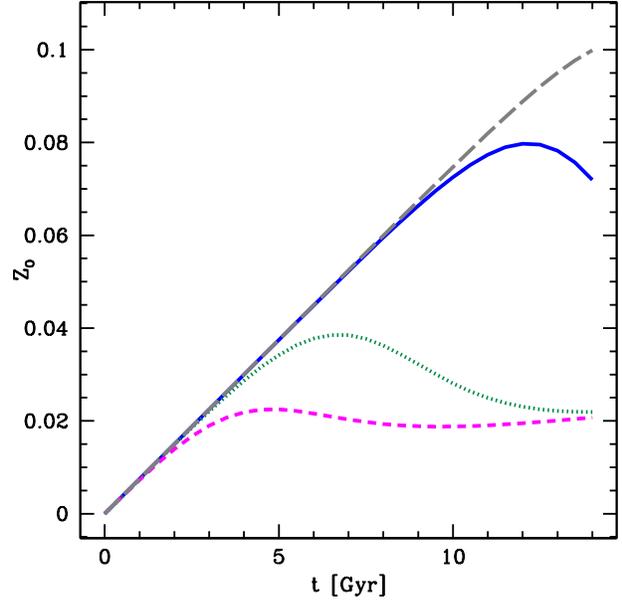} 
  \caption{  We show the time evolution of the metallicity $Z_{O,1}$  when we
    consider different mass ratios  $M_{g1}(0)/ M_{g2}(0)$, whereas   SFEs $S_1$ and $S_2$  have been fixed at the values of 1 and 0.1  Gyr$^{-1}$, respectively. We  use the new
    analytical solution presented in eq. (\ref{sol_SFE_dif}) with  $\epsilon$=0.5, and $\lambda$=0.1. The model with
    $M_{g1}(0)/ M_{g2}(0)$= 10$^3$  is drawn with the grey long solid line, the model
    with   $M_{g1}(0)/ M_{g2}(0)$= 10$^2$   is represented by   the blue
    solid line, the case with $M_{g1}(0)/ M_{g2}(0)$= 1 is drawn with  with the dotted green line. With the magenta short  dashed  line the model with
    $M_{g1}(0)/ M_{g2}(0)$= 10$^{-1}$  is represented. The initial oxygen abundances for the galaxy1 and galaxy2 are  $Z_{O,1}(0)$=$Z_{O,2}(0)$=0.}
\label{2SFEV}
\end{figure}
 This kind of duty cycle is
called galactic fountain (Shapiro \& Field 1976, Bregman
1980). Spitoni et al. (2008, 2009) showed the effect of galactic
fountains on a detailed chemical evolution model (where the
instantaneous recycling and mixing approximation were relaxed) of the
Milky Way. They  discussed the delay in the chemical enrichment due to the
fact that the gas takes a finite time to orbit around the Galaxy an
fall back into the disk. In Recchi et al. (2008) an analytical solution
was presented in the case for galactic fountains.

Here, we want to generalize the results of Recchi et al. (2008), and
 find a new analytical solution for the metallicity (oxygen) in the galaxy1
  when the galactic fountain effect is included together with the
the enriched infall from  the galaxy2. 
To consider galactic fountains we have 
 to take into account an outflow episode and a new infall one
 caused by SN explosion events in the galaxy, with parameters $\omega$
 and $\Omega$. The metallicities of the outflows an fountains are:
\begin{equation}
Z_{\omega}= Z_{O,1} \alpha \omega (1-R) \psi_1,
\end{equation}

\begin{equation}
Z_{\Omega}= Z_{O,1} \alpha \Omega (1-R) \psi_1,
\end{equation}
We are studying galactic fountain effects in the framework of
differential winds as done in Recchi et al. (2008) where it was
introduced the parameter $\alpha>1$ , in order to take into account
for the fact that metals are more easily channeled out from the parent
galaxy compared to the unprocessed gas.

In Fig. \ref{fontfig} we present a
sketch of the gas flow patterns for the galaxy1 when galactic
fountain are considered. It is shown the gas inflow from the galaxy
2 with the associated wind parameter $\lambda$, the outflow from
galaxy1 ($\omega$), and the fraction which comes back
($\Omega$).

Melioli et al. (2015) using three-dimensional hydrodynamical
 simulations investigated the impact of SN feedback in gas rich dwarf
 galaxies and the formation of galactic fountain and outflows.  They
 found a similar circulation flow of the one presented in
 Fig. \ref{fontfig}: galactic fountains is generally established and
 the metal-rich SN ejecta is instead more significant fraction (25-80
 percent) is vented in the intergalactic medium, even.

 It must be underlined that the treatment of the galactic
fountains is actually analogous to what is done in Recchi et
al. (2008) with $\omega$ replacing $\lambda$ and $\Omega$ replacing
$\Lambda$. The difference is in the role of galaxy2.

The system of equations we have to solve for the galaxy1 is the following:
\begin{equation}
\cases{\Scale[1.3]{{d M_{tot1} \over d t}} =  (1 - R)\big( \epsilon \lambda \psi_2 (t)+(\Omega-\omega) \psi_1(t)  \big)  \cr
\Scale[1.3]{{d M_{g1} \over d t}} =   (1 - R)\big( \epsilon \lambda \psi_2 (t)+(\Omega-\omega-1) \psi_1(t)  \big)  \cr
\Scale[1.3]{{d M_{O,1} \over d t}} = (1 - R) \big(\psi_1 (t) \big[ y_O +\Theta  Z_{O,1}(t)\big]+  \psi_2 (t) \epsilon \lambda Z_{O,2}(t)\big).}
\label{system1_fount}
\end{equation}
with $\Theta=- 1 +\alpha\Omega  -\alpha\omega$.

The differential equation of the metallicity in terms of oxygen for the system is:
\begin{displaymath}
\frac{d Z_{O,1}(t)}{dt} = (1 - R) S \times
\end{displaymath}
\begin{equation}
\Big(  y_O   + Z_{O,1}(t) \chi+  \frac{M_{g2}(t)}{M_{g1}(t)} \big[-\epsilon \lambda Z_{O,1}(t) +\epsilon \lambda y_O (1-R)S t   \big]\Big),
\label{}
\end{equation}
with $\chi=(\alpha-1)(\Omega-\omega)$.  It can be easily proved that
the evolution in time of the mass of gas of the galaxy1 is given by:
\begin{displaymath}
M_{g1}(t)= \frac{e^{-(1-\Omega+\omega)(1-R)St}}{1+\Omega-\omega}\times
\end{displaymath}
\begin{equation}
\Big(\lambda(1-e^{-(\lambda+\Omega-\omega)(1-R)St})\epsilon M_{g2}(0)+M_{g1}(0)(\lambda +\Omega-\omega) \Big).
\label{}
\end{equation}
In eq. (\ref{z_fountain}) of the Appendix A we show the complete
expression of the analytical solution in presence of galactic
fountains.

In Fig. \ref{plotfont} the effects of different values of the
parameter $\alpha$ are tested. The higher $\alpha$ is, the less the
galaxy gets enriched. This is because we assumed that $\Omega<\omega$
and consequently the metals (in this case oxygen) which escape, are
larger than the ones which rain back into the galaxy.

\subsection{Model with different star formation efficiencies}
Finally, we test the case where the interacting galaxies have different SFEs.
It is well known that different galactic systems can be characterized
by different SFEs (Matteucci 2001), and generally, higher SFEs can be
associated to more massive systems. 
Here, we show the new analytical solution  for the galaxy1 with a SFE $S_1$   receiving enriched outflow of gas from another galaxy associated with a SFE $S_2$.
 Therefore, with the new SFRs  $\psi_1= S_1 \times M_{g1}$ and $\psi_2= S_2 \times
M_{g2}$  the eq. (\ref{Mg1_SFE_fix}) becomes:

\begin{displaymath}
  M_{g1}(t)=\frac {e^{-( 1-R)  S_1 t}} {  S_1 - ( \lambda + 1)  S_2} \times
   \end{displaymath}

\begin{equation}
 \biggl\lbrace  M_{g1}(0) \Big(  S_1 - ( \lambda + 1)  S_2 \Big)+\epsilon \lambda  M_{g2}(0) S_2\left (e^{\delta t} - 
        1 \right) \biggr\rbrace,
\label{mg1_SFE}
\end{equation}
with $\delta=(1-R) (  S_1 - ( \lambda + 1)  S_2)$. It is trivial to see that when $S_1=S_2$ eq. (\ref{mg1_SFE}) is identical to eq. (\ref{Mg1_SFE_fix}).
The  differential equation we have to solve to have the metallicity evolution of the galaxy1 in presence of different SFEs is:

\begin{displaymath}
\frac{d Z_{O,1}(t)}{dt} = (1 - R)  \times
\end{displaymath}
\begin{equation}
\Big(  y_O S_1 +   S_2\frac{M_{g2}(t)}{M_{g1}(t)} \big[-\epsilon \lambda Z_{O,1}(t) +\epsilon \lambda y_O (1-R)S_2t   \big] \Big).
\label{}
\end{equation}
Finally, the new analytical solution in this case is the following:

\begin{displaymath}
   Z_{O,1}(t)= \frac{1}{(S_1 - (1 + \lambda) S_2)}  \times
   \end{displaymath}

\begin{equation}
 \frac{\Big[\epsilon \lambda   S_2 y_o \Big]A(t) - 
     \frac{M_{g1}(0)}{ M_{g2}(0)} \Big[B(t)\Big] }{ (-1 + 
        e^{\delta t}) \epsilon \lambda    S_2 + 
        \frac{M_{g1}(0)}{ M_{g2}(0)} \Big(S_1 - (1 + \lambda) S_2 \Big)},
\label{sol_SFE_dif}
 \end{equation}
where $A(t)$ and $B(t)$ are the following time dependent terms:
\begin{displaymath}
  A(t)=S_2 + 
      e^{\delta t} \biggl\lbrace S_1 - 
         S_2 + (-1 + R) S_2 \Big(-S_1 + S_2 (1+ \lambda)  \Big) t \biggr\rbrace + 
   \end{displaymath}
\begin{equation}
   + S_1  \biggl\lbrace-1 + (-1 + R) \Big(S_1 - (1 + \lambda) S_2\Big) t \biggr\rbrace,
 \end{equation}

\begin{equation}
  B(t)=(-S_1 + S_2 + \lambda S_2)^2 \Big((-1 + R) S_1 t y_o - 
       Z_{O,1}(0)\Big).
 \end{equation}

We note that the result depends on the ratio between the initial masses of
galaxy1 and galaxy2. Moreover, as expected, when we impose $S_1=S_2$ in
eq. (\ref{sol_SFE_dif}), the resulting $Z_{O,1}(t)$ is identical to
the one obtained with eq. (\ref{znew}). In Fig. \ref{1SFEV} we test
the effect of different SFEs for the galaxy1 and galaxy2 when we
consider the same initial gas  mass ratio ($M_{g1}(0)/ M_{g2}(0)$= 1). We
tested different values for $S_2$: 2, 1, 0.5, 0.1 Gyr$^{-1}$. For the
galaxy1 we assume $S_1$=1 Gyr$^{-1}$ in all models. 

We obtain, as expected, that smaller values of $S_2$ lead to a smaller
chemical enrichment of the system. In the case of $S_2$= 0.1
Gyr$^{-1}$ the metallicity of the system even decreases at later
galactic times due to the strong dilution effect of the infalling
gas. In fact, we can consider it like pristine gas because of, as
expected, the small SFE value of galaxy2 compared to galaxy1.

In Fig. \ref{2SFEV}, fixing $S_1$ and $S_2$ at
the constant values of 1 and 0.1 Gyr$^{-1}$, respectively,  we show the effect of different
initial gas mass ratios for the galaxy1 and galaxy2.  We consider four cases:
$M_{g1}(0)/ M_{g2}(0)$= 10$^3$, 10$^2$, 1, and
10$^{-1}$. Because of the choice of an extremely low SFE for  galaxy2
we expect that the most important deviation from the closed-box
solution is obtained with larger $M_{g2}(0)$ values, hence smaller
$M_{g1}(0)/ M_{g2}(0)$ ratios. In Fig.  \ref{2SFEV} we confirm it, and the model with 
$M_{g1}(0)/ M_{g2}(0)$= 10$^3$, is almost identical to the closed-box evolution model.
\begin{figure}
	  \centering \includegraphics[scale=0.42]{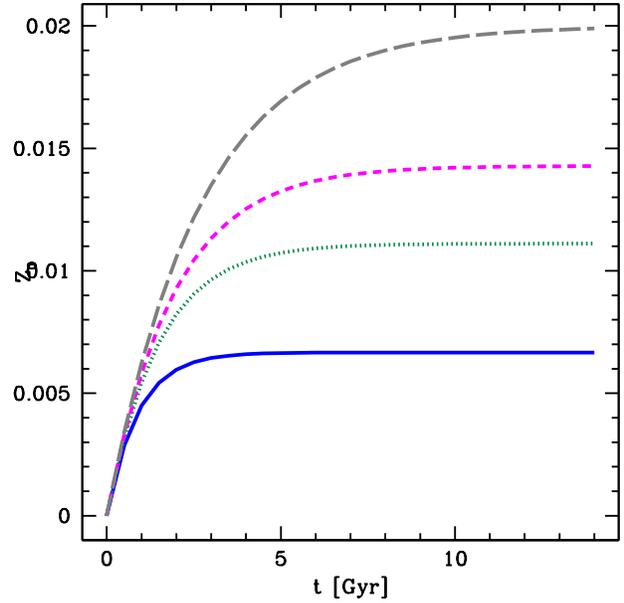} \caption{
    We report the time evolution of the metallicity $Z_{O,1}$ when we
     take into account also  a primordial infall with wind parameter
    $\Lambda$.  We consider the case when $M_{g1}(0)/ M_{g2}(0)$=1,
    $\epsilon$=0.5, $\lambda$=0.3, and  $S$=1 Gyr$^{-1}$.  The initial oxygen abundances for the galaxy1 and galaxy2 are  $Z_{O,1}(0)$=$Z_{O,2}(0)$=0. The model
    with $\Lambda$=0.5 is drawn with the long dashed grey line, the
    model with $\Lambda$=0.7 is represented by the magenta short
    dashed line, the case with $\Lambda$=0.9 is drawn with the dotted
    green line. With the solid blue line the model with $\Lambda$=1.5
    is represented.}
		\label{2infall_constant}
\end{figure}

\begin{figure}
	  \centering \includegraphics[scale=0.42]{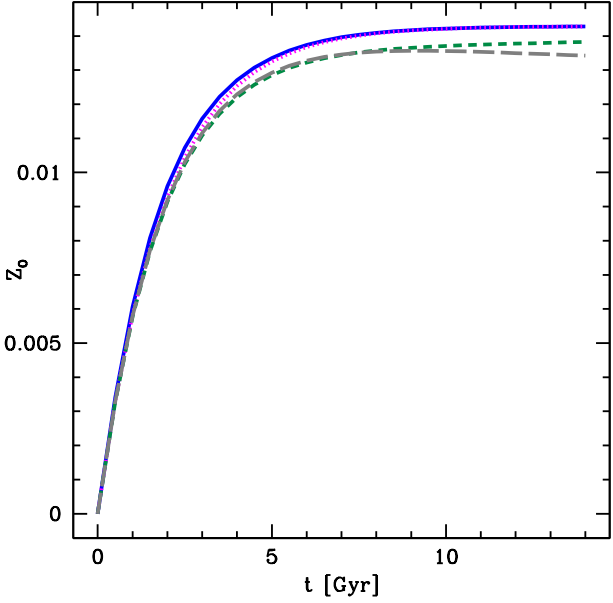} \caption{
    We show the time evolution of the metallicity $Z_{O,1}$ when we
    consider different SFEs $S_2$ for the galaxy2 coupled with a
    primordial infall of gas following the new analytical solution
    presented in eq. (\ref{SFEprim}) of Appendix A. We consider the case
    when $M_{g1}(0)/ M_{g2}(0)$=1, and $\epsilon$=0.5, and the SFE fro
    the galaxy1 is fixed at the value of $S_1$=1 Gyr$^{-1}$ . All the
    models assume $\lambda$=0.3 and $\Lambda$=0.7.  The initial oxygen abundances for the galaxy1 and galaxy2 are  $Z_{O,1}(0)$=$Z_{O,2}(0)$=0. The model with
    $S_2$=2 Gyr$^{-1}$ is represented by the magenta dotted line, the
    model with $S_2$=1 Gyr$^{-1}$ is drawn with the blue solid line,
    the model with $S_2$=0.2 Gyr$^{-1}$ is drawn with
     the green short dashed line.  Finally,
    the model with $S_2$=0.1 Gyr$^{-1}$ is drawn with the grey long
    dashed line.}
		\label{infall_prim2}
\end{figure}

\subsection{The effects of the primordial infall}
 We present here the results  in the presence of a primordial infall of gas for both galaxy1 and galaxy2.  First, we need to compute the time evolution
for the metallicity $Z_{O,2}(t)$ when a primordial infall of gas is
considered.
  To this aim we need the
expression of the mass fraction $\mu_2(t)$ as a function of time. From
the system (\ref{sprimordial2}), with the condition that the total
initial mass is equal to the gas content $M_{tot2}(0)=M_{g2}(0)$ we
have that:
\begin{equation}
M_{tot2}(t)=\frac{\Lambda-\lambda}{\Lambda-\lambda-1}M_{g2}(t)-M_{g2}(0)\frac{1}{\Lambda-\lambda-1}.
\end{equation}
The mass fraction $\mu_2=M_{g2}(t)/M_{tot2}(t) $ can be written as:
\begin{equation}
\mu_2^{-1}=\left(\frac{\Lambda-\lambda}{\Lambda-\lambda-1}-\frac{1}{\Lambda-\lambda-1}e^{-(\Lambda-\lambda -1) (1 - R)St}\right).
\label{muprim}
 \end{equation} We recall that the general solution as  function of
the gas fraction for a system with primordial infall and outflow is
given by eq. (\ref{eq:sol}) with $Z_A=0$. Finally, if we insert
eq. (\ref{muprim}) in eq. (\ref{eq:sol}), we obtain that

\begin{equation}
Z_{O,2} = \frac{ y_O} {\Lambda}
 \left(1-  e^{-\Lambda (1 - R)St} \right).
\label{Z2prim}
\end{equation}
The new expressions for the time evolution of  $M_{g1}(t)$ and $M_{g2}(t)$ are, respectively:

\begin{equation}
 M_{g1}(t) = e^{(\Lambda-1)(1 - R)St}\biggl\lbrace M_{g1}(0)+ \epsilon M_{g2}(0)\big[1- e^{-\lambda(1 - R)St}\big] \biggr\rbrace,
\label{}
\end{equation}

\begin{equation}
 M_{g2}(t) =  M_{g2}(0)e^{(\Lambda-\lambda  -1) (1 - R)St}.
\end{equation}

The differential equation we should to solve in presence of a primordial infall, is the following one:
\begin{displaymath}
\frac{d Z_{O,1}(t)}{dt} = (1 - R) S\big(y_O- \Lambda Z_{O,1}(t)\big)+
\end{displaymath}
\begin{equation}
(1-R) S \frac{M_{g2}(t)}{M_{g1}(t)} \biggl\lbrace \lambda \epsilon  \frac{y_O}{\Lambda} \left[1-  e^{-\Lambda (1 - R)St} \right] - \epsilon Z_{O,1}(t) \lambda  \biggr\rbrace.
\label{Z_prim1}
\end{equation}
The solution of eq. (\ref{Z_prim1}) for the time evolution of the
oxygen abundance for the galaxy1 is then  given by:

\begin{displaymath}
Z_{O,1}(t)=\frac{1}{\Lambda \Big(M_{g1}(0) + (1 - e^{
        \lambda (-1 + R) S t}) \epsilon M_{g2}(0) \Big)} \times
\end{displaymath}

\begin{equation}
 \Big( A2(t) + 
     \big[M_{g1}(0) + \epsilon M_{g2}(0)\big] y_o - 
    B2(t)+C2(t) \Big),
\end{equation}
with
\begin{displaymath}
 A2(t)= (-1 + e^{\Lambda (-1 + R) S t})  e^{\lambda (-1 + R) S t}\epsilon M_{g2}(0) yo,  
   \end{displaymath}

\begin{displaymath}
  B2(t)=    e^{( \Lambda ) (-1 + R) S t}\epsilon M_{g2}(0) y_o,  
   \end{displaymath}

and
\begin{displaymath}
  C2(t)= - e^{( \Lambda ) (-1 + R) S t}M_{g1}(0) \big[y_o - \Lambda  Z_{O,1}(0)\big]. 
\end{displaymath}
It is easy to show that it is possible to recover the solution of
    eq. (\ref{znew}) presented in Section 4.1 for $\lim_{\Lambda \to
    0} Z_{O,1}(t)$. In Fig. \ref{2infall_constant} we show the time
    evolution of the metallicity $Z_{O,1}$ when $M_{g1}(0)/
    M_{g2}(0)$=1, $\epsilon$=0.5, and the SFE for the galaxy1
    and galaxy2 are fixed at the value  $S$=1 Gyr$^{-1}$ . Assuming for
    all the models $\lambda$=0.3, we test different values for the
    the infall parameter $\Lambda$ which is associated with the
    primordial infall: 0.5, 0.7, 0.9, 1.5.

 We note that the effect of dilution of a primordial gas infall
  overwhelms the enriched infall from the companion galaxy, and the
  time evolution of the metallicity for the galaxy1 deviates
  substantially from the closed-box solution (we have proved in
  previous Sections that the evolution in time of a system with only
  enriched gas from a companion galaxy follows the closed-box
  solution) even with a primordial infall parameter comparable to
  the one of enriched infall.

We also consider the case with different SFEs for the galaxy1 and galaxy2
    in presence of primordial infall. We do not enter into details
    about the procedures and the relations used to recover the new
    analytical solution $Z_{O,1}(t)$ , and we only show the solution
    in eq.(\ref{SFEprim}) of Appendix A.

In Fig.  \ref{infall_prim2} we show the effects of different values
for SFEs of the galaxy2 on the chemical evolution of galaxy1 in
presence of primordial infall. We consider the case when the mass
ratio is $M_{g1}(0)/ M_{g2}(0)$=1, and $\epsilon$=0.5, and the SFE for
the galaxy1 is fixed at the value of $S_1=1$ Gyr$^{-1}$ . All the
models assume $\lambda$=0.3 and $\Lambda$=0.7. We consider different
models with different $S_2$ values:  2,1, 0.2, 0.1 Gyr$^{-1}$.  In
Fig.  \ref{infall_prim2} it is clearly shown that all the models with
different SFEs show more or less the same time evolution for the
$Z_{O,1}$. 

 Comparing the models with $S_2$=0.2 Gyr$^{-1}$ and $S_2$=0.1
Gyr$^{-1}$ we note that at late times, as expected, the chemical
evolution of galaxy1, which suffers the enriched infall from galaxy2
with $S_2$=0.2, shows a higher metallicity. This behavior is inverted
at early times. This is due to the way in which we consider the
primordial infall. In fact as shown in eq. (5), the primordial infall
is assumed to be proportional to the SFR of the galaxy, therefore at
early times the galaxy2 with $S_2$=0.2 Gyr$^{-1}$ suffers a larger
dilution effect than the same system with $S_2$=0.1 Gyr$^{-1}$.  

 We conclude that in presence of primordial infall the
chemically enriched gas coming from galaxy2 has not a big
effect on the chemical evolution of the galaxy1. 

 In this work  we consider only primordial infall from the IGM, whereas in
 Peng \& Maiolino (2014b) it was studied how global environment
 properties (overdensity of galaxies) can modify the metallicity of
 the IGM. The observed strong correlation between over-density and
 metallicity for star-forming satellites suggests that the gas infall
 is getting progressively more metal-enriched in dense regions.

\section{Conclusions}

In this paper we presented new analytical solutions for the evolution of the
metallicity (oxygen) of a galaxy in presence of ``environment'' effects
coupled with galactic fountains, and primordial infall of gas.  The
main results of our study are the following ones:

\begin{itemize}
\item If we consider a linear Schmidt (1959) law for the star
  formation rate we have the same time evolution of the metallicity
  for both the closed-box model and the one with only
  outflow. Therefore, in the last case the result does not depend on
  the wind parameter $\lambda$. This result is holding when the
  outflow is not differential. It is worth to note that even if at the
  same time those two systems show the same content of metals they
  show different gas fractions as shown in Fig. \ref{wt}.

\item The new analytical solution for the evolution of the galaxy1 where we consider the enriched inflow of gas from the galaxy2 with  initial metallicities   $Z_{O,1}(0)$ (galaxy1) and   $Z_{O,2}(0)$ (galaxy2) is:

\begin{displaymath}
Z_{O,1}(t) = Sy_O(1-R)t+
\label{znew_gen}
\end{displaymath}
\begin{displaymath}
+\frac{ Z_{O,1}(0)+ Z_{O,2}(0)\epsilon \frac{M_{g2}(0)}{M_{g1}(0)}\Big(1 - e^{-\lambda (1-R)St}\Big) }{1+  \epsilon \frac{M_{g2}(0)}{M_{g1}(0)}\Big(1 - e^{-\lambda (1 - R)St}\Big) }. 
\end{displaymath}
where: $\lambda$ is the wind parameter, $\epsilon$ is the fraction of the outflowing gas   from galaxy2 to galaxy1, $S$ is the star formation efficiency,  and $\frac{M_{g2}(0)}{M_{g1}(0)}$ is the ratio between the initial gas masses of the two galaxies.
\item When we consider the time evolution of a galaxy including an enriched infall
 due to the interactions with a nearby galaxy assuming no pre-enriched
 gas for both galaxies, the chemical enrichemnt is less efficient than
 in the closed-box solution at a fixed stellar mass. Moreover, the
 mass-metallicity relation for galaxies which suffer a gas infall from
 an evolving galaxy, is shifted down at lower Z values if compared to
 the closed-box model results for isolated galaxies.

\item We show a new analytical solution where we consider galactic fountain effects 
in the framework of differential winds coupled with the interaction 
with a nearby galaxy.  
\item We presented the new solution in case of different SFEs. 
If the infall gas originated by a nearby galaxy  has a
  smaller SFE than the galaxy suffering the infall, the recipient
  galaxy will show a less efficient chemical enrichment. In this case
  we showed that the smaller is the ratio between the initial gas  masses
  of the recipient and  donor galaxy, the less efficient the
  chemical evolution is.

\item A new analytical solution when  a primordial infall is coupled with interactions with a nearby galaxy is presented. The effect of dilution of a 
  primordial gas infall overwhelms the enriched infall from the
  chemical evolution of a companion galaxy even with a primordial
  infall parameter comparable to the one of enriched infall.   We
  have also shown how different SFEs for the companion galaxy does not
  affect the chemical evolution of galaxy1 in presence of primordial
  infall.

\end{itemize}

\section*{Acknowledgments}

 The author thanks F. Vincenzo, and  V. Grieco for several fruitful
discussions, and F. Matteucci for many useful suggestions and for
reading the manuscript. The author also thanks an anonymous referee whose
comments improved noticeably the paper.  The work was supported by PRIN
MIUR 2010-2011, project ``The Chemical and dynamical Evolution of the
Milky Way and Local Group Galaxies'', prot. 2010LY5N2T.

\onecolumn
\appendix

\section{New analytical solutions with galactic fountains and primordial infall}

\begin{itemize}
\item
We report here the analytical solution for the time evolution of the
oxygen abundance $Z_{O,1}$ when galactic fountain and the enriched
infall of gas from a companion galaxy are taken into account. This
solution is obtained when the initial metallicity of the galaxy1 is
assumed equal to zero.

\begin{equation}
Z_{O,1}=\frac{e^{-(-1 + \alpha) (\Omega - \omega) (-1 + 
      R) S t} yo }{(-1 + \alpha) (\lambda + 
     \alpha (\Omega - 
        \omega))^2  A3(t) (\Omega - \omega)} \times 
\label{z_fountain}
\end{equation}

\begin{displaymath}
 \Big(\alpha^2 (1 - 
        e^{(-1 + \alpha) (\Omega - \omega) (-1 + R) S t}) M_{g1}(0) (\Omega - 
        \omega)^3 + 
     \lambda^3 B3(t) + 
     \lambda^2 (\Omega - 
        \omega) C3(t) + 
     \lambda (\Omega - \omega)^2 D3(t) \Big)
\end{displaymath}
with

\begin{displaymath}
A3(t)=\lambda (M_{g1}(0) + (1 - 
           e^{(\lambda + \Omega - \omega) (-1 + R) S t}) \epsilon M_{g2}(0)) + 
     M_{g1}(0) (\Omega - \omega)
\end{displaymath}

\begin{displaymath}
B3(t)=\Big(1 - e^{(-1 + \alpha) (\Omega - \omega) (-1 + R) S t}\Big) (M_{g1}(0) + 
           \epsilon M_{g2}(0)) + (-1 + 
           \alpha) e^{(\lambda + \alpha (\Omega - \omega)) (-1 + R) S t}
          \epsilon M_{g2}(0) (\Omega - \omega) (-1 + R) S t
\end{displaymath}

\begin{displaymath}
C3(t)=\Big(1 - e^{(-1 + \alpha) (\Omega - \omega) (-1 + R) S t}\Big) \Big(M_{g1}(0) + 
           2 \alpha M_{g1}(0) + 2 \alpha \epsilon M_{g2}(0) \Big) + (-1 + 
           \alpha^2) e^{(\lambda + \alpha (\Omega - \omega)) (-1 + R) S t}
          \epsilon M_{g2}(0) (\Omega - \omega) (-1 + R) S t,
\end{displaymath}

\begin{displaymath}
D3(t)=-\epsilon M_{g2}(0) + 
        \alpha E3(t) + (-1 +
            \alpha) e^{(\lambda + \alpha (\Omega - \omega)) (-1 + R) S t}
          \epsilon M_{g2}(0) \Big(-1 + \alpha + 
           \alpha (\Omega - \omega) (-1 + R) S t\Big),
\end{displaymath}

\begin{displaymath}
E3(t)=(2 + \alpha) M_{g1}(0) + 2 \epsilon M_{g2}(0) - 
           e^{(-1 + \alpha) (\Omega - \omega) (-1 + 
               R) S t} \Big((2 + \alpha) M_{g1}(0) + \alpha \epsilon M_{g2}(0)\Big).
\end{displaymath}
\item The following solution is related to the case with a primordial infall
coupled with an enriched infall from a companion galaxy, considering
different SFEs:

\begin{equation}
Z_{O,1}(t)=\frac{A4(t) +B4(t)} {\Lambda (S_1 - (1 + 
        \lambda) S_2)(1 - 
         e^{q t}) \epsilon \lambda M_{g2}(0) S_2 + 
    M_{g1}(0) ((-1 + \Lambda) S_1 + (1 - \Lambda + \lambda) S_2)}
\label{SFEprim}
\end{equation}
with:
\begin{displaymath}
q=(-1 + R) \Big((-1 + \Lambda) S_1 + (1 - \Lambda + 
               \lambda) S_2\Big),
\end{displaymath}

\begin{displaymath}
q_2=q/(-1+R),
\end{displaymath}

\begin{displaymath}
\Lambda_*=  \Lambda (-1 + R) S_1,
\end{displaymath}

\begin{displaymath}
\epsilon \lambda M_{g2}(0) S_2,
\end{displaymath}

\begin{displaymath}
A4(t)=\epsilon \lambda M_{g2}(0) S_2 \Big([1 - 
         e^{q t} + (e^{
            \Lambda* t} - 
            e^{q t}) (-1 + 
            \Lambda)] S_1 - [1 + (e^{\Lambda_* t} - 
            e^{q t}) (-1 + \Lambda - 
            \lambda) + \lambda - 
         e^{q t} (1 + 
            \lambda)] S_2 \Big) yo,
\end{displaymath}

\begin{displaymath}
B4(t)=M_{g1}(0) \Big(S_1 - (1 + \lambda) S_2 \Big) q_2 \Big(yo + 
      e^{\Lambda_* t} [-yo + \Lambda  Z_{O,1}(0)]\Big).
\end{displaymath}
\end{itemize}
%\clearpage

\section{List of variables and parameters}

  $Z_{O,i}$: oxygen abundance of galaxy $i$;  \\
 $M_{gi}$:  mass of gas in galaxy $i$; \\
$\psi_i$: star formation rate fo galaxy $i$;\\
$S_i$: star formation efficiency of galaxy $i$;\\
$\mu_i$: gas fraction of galaxy $i$;\\
$y_O$: oxygen yield;\\
$R$: returned fraction;\\
$\lambda$:  outflow parameter;\\
$\Lambda$:  infall parameter;\\
$\epsilon$: fraction of the outflowing gas from galaxy2 to galaxy1;\\
$\omega$: outflow parameter connected to galactic fountains;\\
$\Omega$: infall parameter connected to galactic fountains;\\
$\alpha$: differential wind parameter.\\


\begin{thebibliography}{99}

\bibitem[\protect\citeauthoryear{}{}]{}
Barnes, J. E., Hernquist, L. E. 1992, ARA\&A, 30, 705

\bibitem[\protect\citeauthoryear{}{}]{}
Berentzen, I., Athanassoula, E., Heller, C. H.,  Fricke, K. J. 2003, MNRAS,
341, 343
\bibitem[\protect\citeauthoryear{}{}]{}
Boomsma, R., Oosterloo, T., Fraternali, F., van der Hulst, T.,  Sancisi, R. 2005,
in Extra-planar Gas Conference, ASP Conf. Ser., ed. R. Braun, 331, 247

\bibitem[\protect\citeauthoryear{}{}]{}
Boselli, A., \& Gavazzi, G.  2006, PASP, 118, 517 
\bibitem[\protect\citeauthoryear{}{}]{}
Bregman, J. N. 1980, ApJ, 365, 544



\bibitem[\protect\citeauthoryear{}{}]{}
Clayton, D. D. 1988, MNRAS, 234, 1


\bibitem[\protect\citeauthoryear{}{}]{}
 Davies, J.~I., Wilson, 
C.~D., Auld, R., et al., 2010, MNRAS, 409, 102 

\bibitem[\protect\citeauthoryear{}{}]{}
Edmunds, M. G., 1990, MNRAS, 246, 678

\bibitem[\protect\citeauthoryear{}{}]{}
Fraternali, F., Oosterloo, T., Sancisi, R. 2004, A\&A, 424, 485

\bibitem[\protect\citeauthoryear{}{}]{}
Greggio, L.,  Renzini, A. 1983, A\&A, 118, 217



\bibitem[\protect\citeauthoryear{}{}]{}
Hartwick, F. D. A., 1976, ApJ, 209, 418

\bibitem[\protect\citeauthoryear{}{}]{}
Hoopes, C.G., Heckman, 
T.M., Strickland, D.K., et al. 2005, ApJL, 619, L99 



\bibitem[\protect\citeauthoryear{}{}]{}
Houck, J. C., Bregman, J. N., 1990, ApJ, 352, 506


\bibitem[\protect\citeauthoryear{}{}]{}
Kennicutt, R. C. 1990, in IAU Colloquium 124, NASA Marshall Space Flight
Center, Paired and Interacting Galaxies, ed. J. W. Sulentic, W. C. Keel, \&
C. M. Telesco, 269

\bibitem[\protect\citeauthoryear{}{}]{}
Kewley, L. J.,  Ellison, S. L. 2008, ApJ, 681, 1183

\bibitem[\protect\citeauthoryear{}{}]{}
Kudritzki, R.P., Ho, 
I., Schruba, A., et al., 2015, arXiv:1503.01503 



\bibitem[\protect\citeauthoryear{}{}]{}
Lacey, C. G.,  Fall, M. 1985, ApJ, 290, 154



\bibitem[\protect\citeauthoryear{}{}]{}
Larson, R. B.,  Tinsley, B. 1978, ApJ, 219, 46

\bibitem[\protect\citeauthoryear{}{}]{}
Lilly S. J., Carollo C. M., Pipino A., Renzini A., Peng Y., 2013, ApJ, 772,
19 

\bibitem[\protect\citeauthoryear{}{}]{}
Marconi, G., Matteucci, F.,  Tosi, M. 1994, MNRAS, 270, 35


\bibitem[\protect\citeauthoryear{}{}]{}
Matteucci, F. 2001, The Chemical Evolution of the Galaxy, ASSL, Kluwer
Academic Publisher


\bibitem[\protect\citeauthoryear{}{}]{}
Matteucci, F.,  Chiosi, C., 1983, A\&A, 123, 121

\bibitem[\protect\citeauthoryear{}{}]{}
Matteucci, F.,  Greggio, L. 1986, A\&A, 154, 279



\bibitem[\protect\citeauthoryear{}{}]{}
Melioli, C., Brighenti, F.,  D'Ercole, A., 2015, MNRAS, 446, 299 

\bibitem[\protect\citeauthoryear{}{}]{}
Mo, H. J., Mao, S.,  White, S. D. M., 1998, MNRAS, 295, 319


\bibitem[\protect\citeauthoryear{}{}]{}
	Pipino, A., Lilly, S. J., Carollo, C. M.,2014, MNRAS, 441, 1444

\bibitem[\protect\citeauthoryear{}{}]{}
Pagel, B. E. J.,  Patchett, B. E. 1975, MNRAS, 172, 13



\bibitem[\protect\citeauthoryear{}{}]{}
Peeples M. S., Shankar F., 2011, MNRAS, 417, 2962

\bibitem[\protect\citeauthoryear{}{}]{}
Peng, Y.,  Maiolino, R. 2014a, MNRAS, 443, 3643


\bibitem[\protect\citeauthoryear{}{}]{}
Peng, Y.,  Maiolino, R. 2014b, MNRAS, 438, 262

\bibitem[\protect\citeauthoryear{}{}]{}
Pilyugin, L. S. 1993, A\&A, 277, 42


\bibitem[\protect\citeauthoryear{}{}]{} 
Recchi, S., Kroupa, P, MNRAS, 2015, 446, 4168 


\bibitem[\protect\citeauthoryear{}{}]{} 
Recchi, S., Kroupa, P, Ploeckinger, S.,  2015,  arXiv:1504.02473 


\bibitem[\protect\citeauthoryear{}{}]{} 
Recchi, S., Spitoni, E., Matteucci, F., Lanfranchi, G. A., 2008, A\&A, 489, 555



\bibitem[\protect\citeauthoryear{}{}]{}
Salpeter, E. E. 1955, ApJ, 121, 161



\bibitem[\protect\citeauthoryear{}{}]{}
Schmidt, M. 1963, ApJ, 137, 758

\bibitem[\protect\citeauthoryear{}{}]{}
Schmidt, M. 1959, ApJ, 129, 243

\bibitem[\protect\citeauthoryear{}{}]{}
Searle, L.,  Sargent, W. L. W. 1972, ApJ, 173, 25
\bibitem[\protect\citeauthoryear{}{}]{}
Shapiro, P. R.,  Field, G. B. 1976, ApJ 205, 762

\bibitem[\protect\citeauthoryear{}{}]{}
Silk, J., 2003, MNRAS, 343, 249 

\bibitem[\protect\citeauthoryear{}{}]{}
Smith, B. J., Giroux, M. L., Struck, C., Hancock, M.,  Hurlock, S. 2010, AJ,
139, 1212

\bibitem[\protect\citeauthoryear{}{}]{}
Smith, B. J., Struck, C., 2001, AJ, 121, 710






\bibitem[\protect\citeauthoryear{}{}]{}
Spitoni E., Calura F., Matteucci F., Recchi S., 2010, A\&A, 514, A73

\bibitem[\protect\citeauthoryear{}{}]{}
Spitoni, E., Recchi, S.,  Matteucci, F. 2008, A\&A, 484, 743

\bibitem[\protect\citeauthoryear{}{}]{}
Spitoni, E.,  Matteucci, F.,Recchi, S., Cescutti, G., Pipino, A., 2009, A\&A, 504, 87



\bibitem[\protect\citeauthoryear{}{}]{}
Temporin, S., Weinberger, R., Galaz, G.,  Kerber, F. 2003a, ApJ, 584, 239
\bibitem[\protect\citeauthoryear{}{}]{}
Temporin, S., Weinberger, R., Galaz, G.,  Kerber, F. 2003b, ApJ, 587, 660



\bibitem[\protect\citeauthoryear{}{}]{}
Tinsley, B. M. 1974, ApJ, 192, 629

\bibitem[\protect\citeauthoryear{}{}]{}
Tinsley, B. M. 1980, Fund. Cosmic Phys., 5, 287
\bibitem[\protect\citeauthoryear{}{}]{}
Toomre, A.,  Toomre, J. 1972, ApJ, 178, 623

\bibitem[\protect\citeauthoryear{}{}]{}
Torrey P., Cox T. J., Kewley L., Hernquist L., 2012, ApJ, 746, 108

\bibitem[\protect\citeauthoryear{}{}]{}
Twarog, B. A., 1980, ApJ, 242, 242
\bibitem[\protect\citeauthoryear{}{}]{}
Woosley, S.E., Weaver, T.A. 1995, ApJS, 101, 181


\end{thebibliography}
\end{document}